\documentclass[aps,prr,preprint,superscriptaddress]{revtex4-2}
\usepackage{graphicx}
\usepackage{float}
\usepackage{bm}
\usepackage[export]{adjustbox}
\usepackage{color}

\newcommand	{\ssf}		{\sigma_s^f}
\newcommand	{\sspfp}	{\sigma_{s^\prime}^{f^\prime}}

\newcommand	{\NHH}	{N_{\text{HH}}}
\newcommand	{\EHP}	{E_{\text{HP}}}
\newcommand  {\lv}           {\bm{\lambda}}

\begin{document}

% Use the \preprint command to place your local institutional report
% number in the upper righthand corner of the title page in preprint mode.
% Multiple \preprint commands are allowed.
% Use the 'preprintnumbers' class option to override journal defaults
% to display numbers if necessary
%\preprint{}

%Title of paper
\title{Folding lattice proteins with quantum annealing}

% repeat the \author .. \affiliation  etc. as needed
% \email, \thanks, \homepage, \altaffiliation all apply to the current
% author. Explanatory text should go in the []'s, actual e-mail
% address or url should go in the {}'s for \email and \homepage.
% Please use the appropriate macro foreach each type of information

% \affiliation command applies to all authors since the last
% \affiliation command. The \affiliation command should follow the
% other information
% \affiliation can be followed by \email, \homepage, \thanks as well.

\author{Anders Irb\"ack}
\email[]{anders.irback@thep.lu.se}
\affiliation{Computational Biology \& Biological Physics, Department of Astronomy,
and Theoretical Physics, Lund University, Box 43, SE-221 00 Lund, Sweden}

%\homepage[]{Your web page}
%\thanks{}
%\altaffiliation{}

\author{Lucas Knuthson}
\affiliation{Computational Biology \& Biological Physics, Department of Astronomy,
and Theoretical Physics, Lund University, Box 43, SE-221 00 Lund, Sweden}

\author{Sandipan Mohanty}
\affiliation{Institute for Advanced Simulation, J\"ulich
Supercomputing Centre, Forschungszentrum J\"ulich, D-52425 J\"ulich, Germany}

\author{Carsten Peterson}
\affiliation{Computational Biology \& Biological Physics, Department of Astronomy,
and Theoretical Physics, Lund University, Box 43, SE-221 00 Lund, Sweden}

%Collaboration name if desired (requires use of superscriptaddress
%option in \documentclass). \noaffiliation is required (may also be
%used with the \author command).
%\collaboration can be followed by \email, \homepage, \thanks as well.
%\collaboration{}
%\noaffiliation

\date{\today}

\begin{abstract}
Quantum annealing is a promising approach for obtaining good approximate solutions to difficult
optimization problems. Folding a protein sequence into its minimum-energy structure represents such a problem.
For testing new algorithms and technologies for this task, the minimal lattice-based HP model is well suited,
as it represents a considerable challenge despite its simplicity.
The HP model has favorable interactions between adjacent, not directly bound hydrophobic residues. 
Here, we develop a novel spin representation for
lattice protein folding tailored for quantum annealing. With a distributed encoding onto the lattice,
it differs from earlier attempts to fold lattice proteins on quantum annealers, which were based upon chain
growth techniques. With our encoding, the Hamiltonian by design has the   
quadratic structure required for calculations on an Ising-type annealer, without having to introduce any
auxiliary spin variables. This property greatly facilitates the study of long chains. The approach 
is robust to changes in the parameters required to constrain the spin system to chain-like configurations, 
and performs very well in terms of solution quality.   
%As a result, our approach naturally implements self-avoidance, performs very well in terms of
%solution quality, and has better scaling properties with chain length. Moreover, the encoding is robust
%to changes in the parameters required to constrain the spin system to chain-like configurations.
The results are evaluated against existing exact results for HP chains with up to $N=30$ beads with 100\% hit rate,
thereby also outperforming classical simulated annealing. In addition, the method allows us to recover the lowest known
energies for $N=48$ and $N=64$ HP chains, with similar hit rates. These results are obtained by the commonly used hybrid quantum-classical approach.
For pure quantum annealing, our method successfully folds an $N=14$ HP chain.
The calculations were performed on a D-Wave Advantage quantum annealer.
\end{abstract}

% insert suggested keywords - APS authors don't need to do this
%\keywords{}

%\maketitle must follow title, authors, abstract, and keywords
\maketitle

% body of paper here - Use proper section commands
% References should be done using the \cite, \ref, and \label commands

\newpage

\section{Introduction\label{sec:intro}}

Quantum computers with their interacting qubits as basic units appear very promising for optimization problems with binary
variables such as those present in spin systems. These technologies are being developed along two main tracks:
quantum annealers~\cite{Johnson:11} and gate-based systems~\cite{Arute:19}.
In quantum annealing (QA)~\cite{Kadowaki:98,Brooke:99,Boixo:14}, the idea is to encode the solution to a given
optimization problem in the ground state of a Hamiltonian and efficiently locate
the energy minimum by exploiting quantum fluctuations and tunneling, in analogy with the
role played by thermal fluctuations in classical simulated annealing (SA)~\cite{Kirkpatrick:83}.
Mapping difficult binary optimization problems onto Ising spin glass systems goes back to the eighties in the context of
neural networks~\cite{Hopfield:85,Peterson:88}. For a recent review, see Ref.~\cite{Lucas:14}.
In the context of QA, this approach is called quadratic unconstrained binary optimization (QUBO).

Protein folding, going from sequence to structure by minimizing an energy function, represents a difficult optimization problem.
Simplified lattice-based models for this problem can often provide qualitatively relevant results, but remain computationally challenging and
are therefore ideal testbeds for novel algorithms. A pioneering QUBO formulation of the folding problem for lattice proteins was given by
Perdomo and coworkers~\cite{Perdomo:08}. They considered the HP model~\cite{Lau:89}, where proteins are represented by linear
chains of $N$ hydrophobic (H) or polar (P) beads, residing on a lattice. Their model used binary variables  
encoding bead coordinates on the lattice, and an additional set of auxiliary binary variables had to be added in order to obtain 
a quadratic Hamiltonian.  
%which led to a relatively large number of required qubits even for short chains but a polynomial rather than exponential scaling with $N$.

An early attempt to fold a short lattice protein ($N=6$) by QA was carried out on a D-Wave (D-Wave Systems Inc.)
machine~\cite{Perdomo-Ortiz:12}. This implementation relied on a growth algorithm, where turns along the chain were mapped onto qubits.
%This encoding is resource-efficient for very short chains, but scales exponentially with $N$. 
Recent work implemented on an IBM
gate-based quantum computer used a similar encoding~\cite{Robert:21}, again for a short protein chain ($N=7$). The growth
algorithms offer a compact, resource-efficient representation of the structure of a chain, but creating a quadratic Hamiltonian 
requires additional spin variables and implementing interactions such as self-avoidance becomes tedious for long chains
with this type of encoding.   
For a recent review of lattice protein folding on quantum computers, see Ref.~\cite{Outeiral:21}.

In contrast to Refs.~\cite{Perdomo-Ortiz:12,Robert:21}, here we propose a different binary encoding for lattice proteins. 
In our model, all sites of the lattice host qubits. The approach was inspired by recent D-Wave applications for
homopolymers \cite{Micheletti:21}, and shares some similarities with a QUBO formulation for lattice
heteropolymers~\cite{Babbush:14} which, to our knowledge, has not yet been implemented.
The energy $E$ is such that its minimization makes the values of the qubits coalesce to a finite set of active qubits
defining the desired folded structure. Importantly from the viewpoint of QA, the entire function
$E$, including a self-avoidance term, is manifestly quadratic, or two-local, in the binary spin variables,
without having to add any auxiliary spins. As a result, $E$ retains a convenient form for long chains.
This fully distributed dynamical encoding method can be considered as a clustering approach driven by requiring
a legal chain on the lattice. It is also somewhat reminiscent of a molecular field theory \cite{Reva:96} where the fields
reside on a lattice and give rise to structures through their dynamics.

We evaluate the performance of our approach using the two-dimensional HP model~\cite{Lau:89} as a testbed. 
Although this model at first sight looks simple, finding its lowest energy has been shown to belong 
to the class of NP-complete problems \cite{Crescenzi:98}. In particular, we consider
a set of sequences with up to 30 beads, for which the exact solutions are known from exhaustive enumerations of all
structures~\cite{Irback:02,Holzgrafe:11}. In addition, we present results obtained for two longer sequences with 48
and 64 beads, respectively, which have been studied by various classical methods~\cite{Unger:93,Bastolla:98,Liang:01}.

In what follows, we first briefly describe the HP model and then map the energy minimization problem for HP sequences
to a QUBO problem. We then evaluate our encoding using both classical SA
and explorations performed on the D-Wave Advantage quantum annealer, with over 5,000
qubits and 15-way qubit connectivity~\cite{McGeoch:20}.

%%%%%%%%%%%%%%%%%%%%%%%%%%%%%%%%%%%%%%%%%%%%%%%%%%%%%%%%%%%%%%%%%%%%%%

\section{Methods}

\subsection{HP lattice proteins}

We consider the minimal lattice-based HP model for protein folding~\cite{Lau:89}, in which
the protein is represented by a self-avoiding chain of hydrophobic (H) or polar (P) beads
on a lattice. Two beads are said to be
in contact if they are nearest neighbors on the lattice, but not along the chain. A given chain
configuration is assigned an energy defined as $\EHP=-\NHH$, where $\NHH$ is the number of
HH contacts~\cite{Lau:89}. With this choice of energy, low-energy configurations tend
to exhibit a hydrophobic core of H beads. Despite the simplicity of the model, there are
HP sequences with a unique ground state, and therefore a well-defined structure. In particular, on a
2D square lattice, it is known from exhaustive enumerations that about 2\% of all HP sequences
with $\le$30 beads have a unique ground state~\cite{Irback:02,Holzgrafe:11}.

\subsection{Binary quadratic model for HP lattice proteins -- QUBO encoding\label{sec:qubo}}

Given an HP sequence $(h_1,\ldots,h_N)$, $h_i\in\{\text{P},\text{H}\}$, we wish to determine its
ground state using QA. To this end, in this section, we present a binary encoding
for HP lattice proteins, assuming a square
grid with $L^2$ sites.

\begin{figure}
    \centering
    \includegraphics[width=5.2cm]{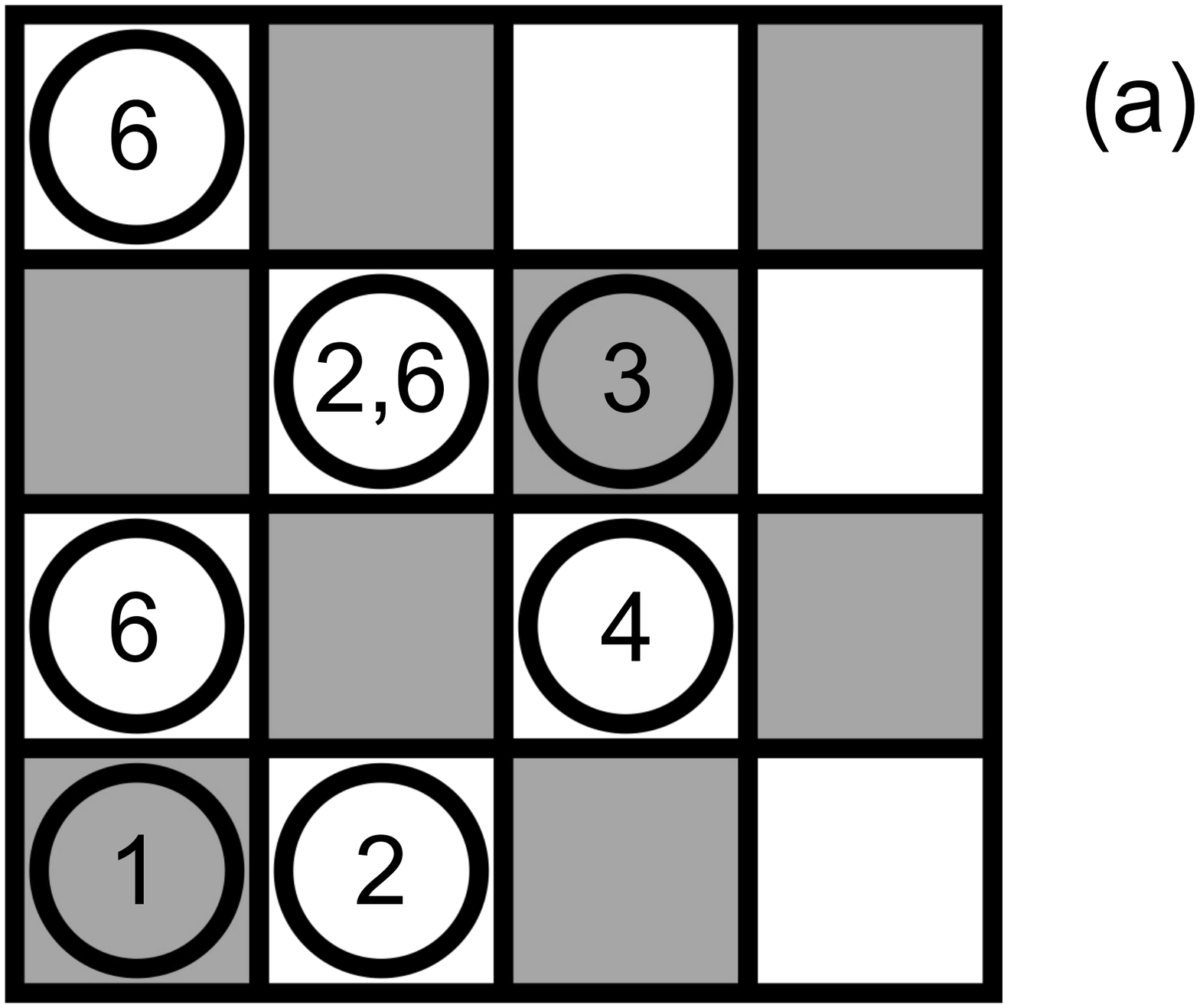}
    \includegraphics[width=5.25cm]{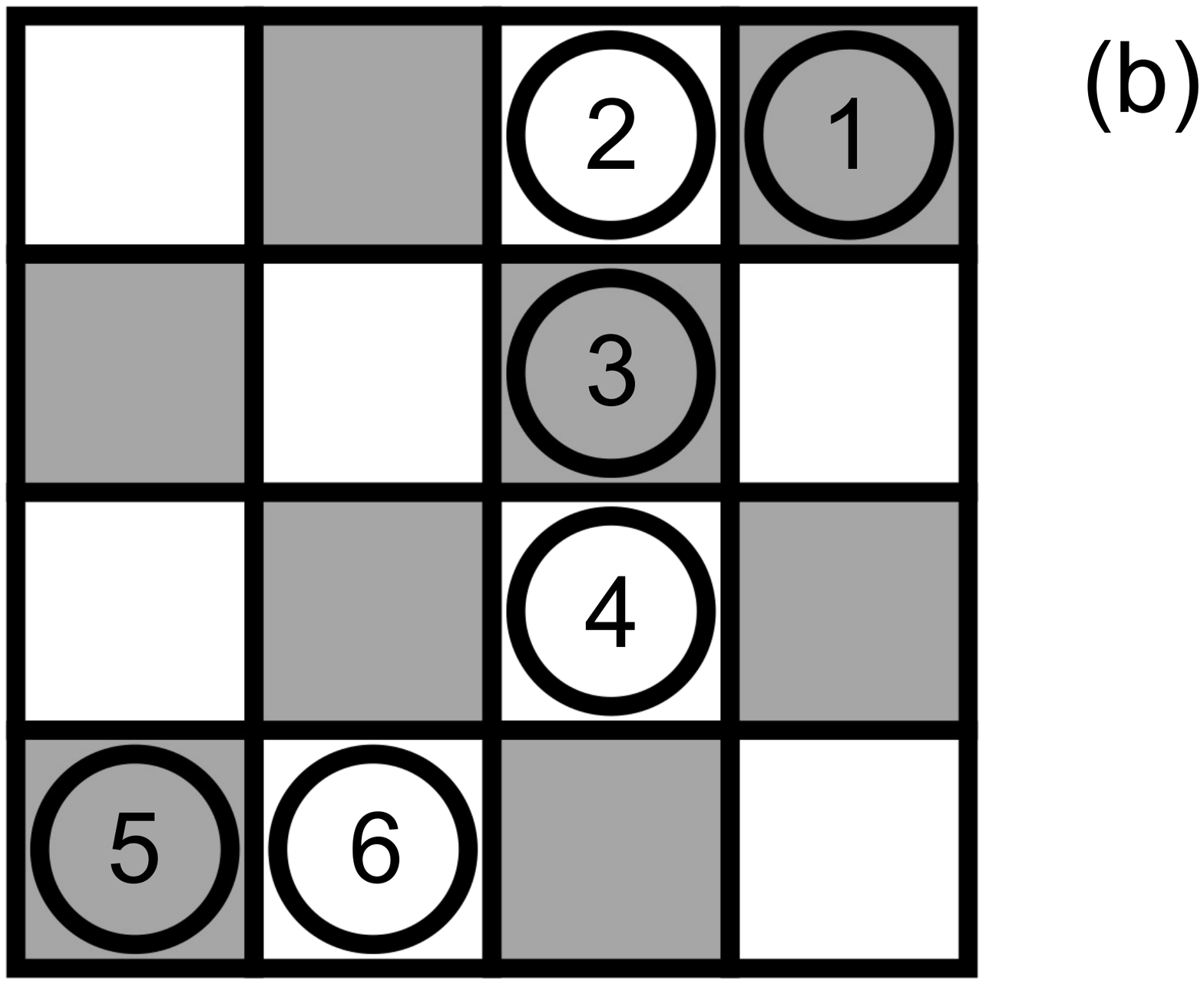}
    \includegraphics[width=5.25cm]{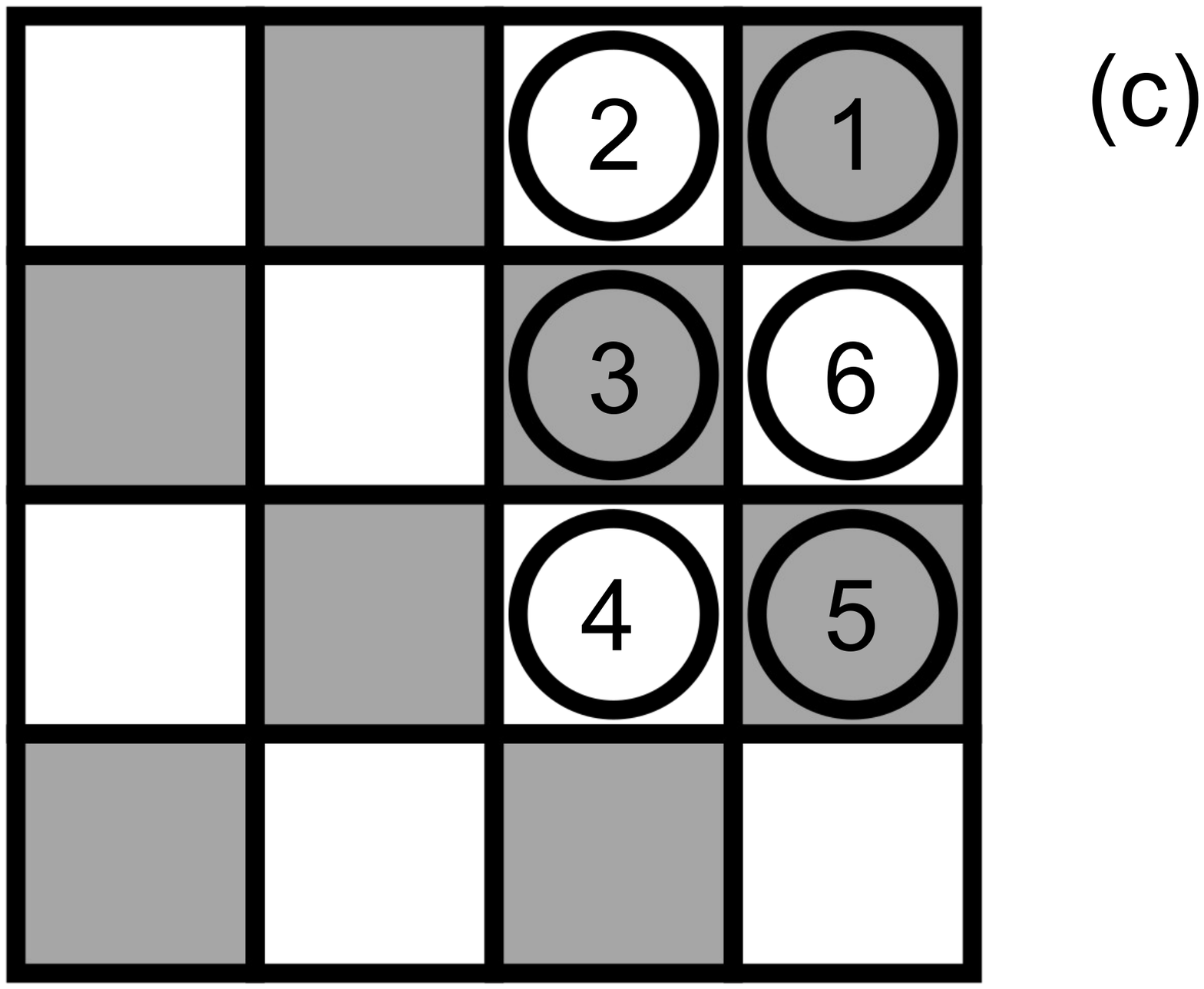}
    \vspace{-4\baselineskip}
    \caption{Illustration of a hypothetical evolution of the binary model described in Sec.~\ref{sec:qubo} for the 6-bead sequence HPHPPH.
    Circles represent beads, and numbers indicate bead positions along the
    sequence. By construction, odd/even beads can reside only on grey/white sites. (a) Early stage. Typically, all the
    three constraints are violated ($E_1, E_2, E_3>0$).  (b) Intermediate stage. Some but not all of the constraints are
    satisfied (in this example: $E_1=E_2=0$, $E_3>0$). (c) The final state, in this example corresponding to the desired
    minimum-energy structure of the given sequence ($\EHP=-2$, $E_1=E_2=E_3=0$).
    \label{fig:method_qubo}}
\end{figure}

Inspired by the binary representation of homopolymers of
Ref.~\cite{Micheletti:21}, rather than directly encoding chain configurations, we introduce fields
of binary variables along with penalty terms. These terms serve to ensure that the final binary field
configurations correspond to proper chain configurations. To reduce the number of binary variables, we
make a checkerboard division of the lattice into even and odd sites, and use the fact that in a valid
chain configuration all even (odd) beads share the same lattice site parity (see Fig.~\ref{fig:method_qubo}). As a result, we may assume
that even (odd) beads reside on even (odd) lattice sites. Thus, we introduce one set of binary fields, $\ssf$, to
describe the location of even beads, and another set for odd beads $\sspfp$. Here, the indices $f$ and
$s$ run over even beads and sites, respectively, while $f^\prime$ and $s^\prime$ run over odd beads
and sites. We set $\ssf=1$ if bead $f$ is located on site $s$, and $\ssf=0$ otherwise. The odd fields
$\sspfp$ are defined in the same way. The division into even and odd sites reduces the number of
variables required from $N\times L^2$ to $\approx$$N\times L^2/2$.

Having defined the degrees of freedom, we now describe the energy function. In our QUBO model,
the total energy $E$ has the form
\begin{equation}\label{eq:E}
E=\EHP+\sum_{i=1}^3 \lambda_i E_i\,,
\end{equation}
where $\EHP$ is the energy of the HP model (see above) and the remaining three terms $E_1$, $E_2$
and $E_3$ are constraint energies.  The strengths of the constraints are set by the parameters $\lambda_i$.

Specifically, in terms of the binary fields, the four energies can be expressed as follows.

\begin{itemize}

\item The HP energy $\EHP=-\NHH$ can be rewritten as
\begin{equation}
\label{eq:EHP}
\EHP=-\sum_{| f-f^\prime |>1}C(h_f,h_{f^\prime})\sum_{\langle s,s^\prime\rangle}\ssf\sspfp
\end{equation}
where the interaction strength $C(h_f,h_{f^\prime})=1$ if $h_f=h_{f^\prime}=\text{H}$ and
$C(h_f,h_{f^\prime})=0$ otherwise. In Eq.~\ref{eq:EHP},
the second sum runs over all nearest-neighbor pairs of sites, $\langle s,s^\prime\rangle$.
Such a pair always
consist of one even and one odd site. The beads $f$ and $f^\prime$ must both be of type H for a non-zero
energy contribution, and must not, with our definition of a contact, be adjacent along the chain.

\item The first constraint energy, $E_1$, is given by
\begin{equation}
E_1= \sum_f\left(\sum_s\ssf-1\right)^2\,+ \{\text{same for odd parity}\}\,,
%\sum_{f^\prime}\left(\sum_{s^\prime}\sspfp-1\right)^2\,,
\end{equation}
and serves to ensure that each bead is located at exactly one lattice site.

\item The energy $E_2$ makes the chain self-avoiding. It is given by
\begin{equation}
E_2=\frac{1}{2}\sum_{f_1\ne f_2} \sum_s \sigma_s^{f_1}\sigma_s^{f_2}+\{\text{same for odd parity}\}\,,
%\frac{1}{2}\sum_{f_1^\prime\ne f_2^\prime} \sum_{s^\prime}\sigma_{s^\prime}^{f_1^\prime}\sigma_{s^\prime}^{f^\prime_2}\,,
\label{eq:H2}
\end{equation}
and provides an energy penalty whenever two beads occupy the same site.

\item The final energy, $E_3$, has the form
\begin{equation}
E_3= \sum_{1\le f<N}\sum_s \sigma_s^f
\sum_{||s^\prime-s||>1}\sigma_{s^\prime}^{f+1}+\{\text{same with odd/even parity interchanged}\}\,,
%\sum_{\substack{s^\prime\ne s \\  s,s^\prime\ \text{non-n.n.}}}\sigma_{s^\prime}^{f+1}\,.
\end{equation}
and is responsible for connecting the beads to a chain. It provides an energy penalty whenever
two adjacent beads along chain are not nearest neighbors on the lattice.

\end{itemize}

Our model contains three parameters; $\lambda_1$, $\lambda_2$ and $\lambda_3$ (Eq.~\ref{eq:E}).
It is desirable that when executing the model it is reasonably robust with respect to these parameters.
This will turn out to be the case in Sec.~\ref{sec:results} when exploring the method.

As indicated in Sec.~\ref{sec:intro}, the above binary model shares similarities
with the ``diamond'' encoding proposed in Ref.~\cite{Babbush:14}.
The latter method is able to reduce the number of binary variables required for very short chains, by
fixing the position of the first bead and using the fact that odd and even beads can be assumed to belong
to different ``diamond'' layers. For long chains, our choice of a freely moving chain on a simple odd/even
checkerboard is more resource-efficient, because, in general, the search for the ground state can be carried
out on a smaller grid if the chain is freely moving. Our constraint energies $E_i$ also differ from those of
Ref.~\cite{Babbush:14}. In our case, all three constraint energies are manifestly non-negative for both
physical and unphysical spin configurations, which makes our method more robust to changes in the strength
parameters $\lambda_i$. Since the encoding in Ref.~\cite{Babbush:14} was never explored, a robustness 
analysis is not available there.

\subsection{Scaling properties}

The binary model introduced above uses $\approx$$NL^2/2$ spins to describe the
structure of an $N$-bead HP protein on an $L^2$ grid. In order that the (in general
\textit{a priori} unknown) minimum-energy structure fits inside the simulation grid,
the lattice size $L$
has to be chosen according to the protein sequence. A safe choice is to take $L=N$,
in which case the
number of spins scales cubically with $N$. However, a typical minimum-energy structure is
compact, since it contains many attractive H-H contacts,
and fits onto a grid with $L$ not much larger than $\sqrt{N}$. Such a structure
may therefore be found using $N^2$
rather than $N^3$ spins. Note that the assumption of a square grid is unnecessary and was only
made for simplicity. For a general grid shape, the number of spins scales as $N$ times the
number of grid points.

Compared to our encoding, turn-based ones~\cite{Perdomo-Ortiz:12,Babbush:14}
provide a more economic description of the chain geometry, requiring only $\sim$$N$ spins.
However, additional spins are needed in order to formulate the Hamiltonian. A
resource-efficient turn-based binary model with a quadratic Hamiltonian 
was proposed in \cite{Babbush:14}, which in total uses $\sim$$N^2$ spins. 
Here, the number of spins is thus comparable to that of our model when using a 
reduced grid size. However, in the turn-based model~\cite{Babbush:14},
the Hamiltonian is somewhat complicated 
already for $N=6$, and it would be challenging to implement it for the
chain lengths studied here. In contrast, the Hamiltonian in
Sec.~\ref{sec:qubo} retains its simple and convenient structure as     
the chain length is increased.\newline

\subsection{Simulated annealing
\label{sec:sa_methods}}

Before turning to QA, we tested this QUBO model using SA,
with the system defined by the partition function $Z=\sum_{\{\ssf,\sigma_{s^\prime}^{f^\prime}\}}e^{-\beta E}$,
where $\beta$ denotes inverse temperature and $E$ is given by Eq.~\ref{eq:E}. All runs spanned
the same set of 25 temperatures, given by $\beta_0=1$ and $\beta_{i+1}=1.05\beta_i$. At each
temperature, $10^4$ sweeps were performed, where one sweep comprises, on average, one
attempted update per spin variable.  The updates were single-spin flips, controlled by a Metropolis
acceptance criterion. All runs were started from random initial spin configurations, and used a
$10^2$ grid.

For comparison, we also conducted SA runs based on the conventional explicit-chain
representation of the HP model. Here, the energy was given by $\EHP$, without the
constraint terms. The set of temperatures was the same as in the QUBO SA runs.
The simulations used three standard Metropolis-type elementary moves: local one- and two-bead
updates, and a non-local pivot update. At each temperature, $10^5$ sweeps were performed,
with one sweep consisting of $N-1$ one-bead moves, $N-2$ two-bead moves and one pivot move.
The chains were not confined to a finite-size grid.

All SA simulations were run on a standard desktop computer. For $N=30$, each QUBO SA run required
21 CPU-core-seconds, whereas each explicit-chain SA run required 5 CPU-core-seconds. To gather
statistics, for each sequence, we performed 1000 runs with each method, using different random
number seeds.

\subsection{Hybrid quantum-classical computations}

D-Wave offers access to solvers based solely on QA as well as hybrid quantum-classical solvers.
The hybrid approach uses classical solvers while sending suitable subproblems as queries to the quantum processing
unit (QPU). The solutions to the subproblems serve to guide the classical solvers \cite{McGeoch:20b}. The goal is to speed up
the solution of challenging QUBO problems by queries to the QPU. With the hybrid approach, it is possible to
tackle much larger problems, with many thousands of fully connected variables, than what can be dealt with
using QA alone.

We conducted hybrid quantum-classical computations for HP chains with up to 64 beads, using D-Wave's
hybrid solver services and a D-Wave Advantage quantum computer.
All sequences with $N\le 30$ were folded on a $10^2$ grid using the default run time set by the hybrid solver,
which depends on problem size and was 4\,s for $N=30$. To gather statistics, a set of 100 runs were performed for each 
sequence. Two additional sequences were studied, with $N=48$ and $N=64$, respectively, using 
both $10^2$ and $15^2$ grids. For these sequences, the run time for the hybrid solver had to be taken larger 
than the default run time, to ensure satisfactory hit rates. For a given sequence and grid, computations  
were performed for a number of different run times, to investigate the run time dependence of the hit rate. For each
choice of sequence, grid and run time, 10 runs were performed.

\subsection{Pure QPU computations\label{sec:QPU_methods}}

The 5000-qubit Advantage machine uses a Pegasus topology with a connectivity of 15~\cite{McGeoch:20}.
Thus, in order to solve a problem with higher connectivity, it has to be embedded into the
Pegasus graph. This embedding is done by forming ``chains'' of qubits that act as single qubits.
The strength of the coupling between the qubits within a chain is a tunable parameter, called the chain strength.
This parameter is typically chosen slightly larger than the minimum chain strength needed to avoid chain breaks.

D-Wave offers several so-called samplers for finding embeddings into the QPU topology and performing
the QPU computation. We used the \texttt{DWaveCliqueSampler},
designed for dense binary quadratic models. All the computations used a chain strength between 1
of 7.5 and the annealing time was set to $\tau=2000\,\mu$s, its maximum allowed value. The number
of output reads (annealing cycles) per run, which must be $<$$10^6/(\tau/\mu$s), was set to 490. 
For each system studied, 100 runs were executed, each with this number of annealing cycles.

\subsection{Testbed -- HP sequences}

As a testbed, we use a selected set of HP sequences with 4--30 beads, all of which are known
from exhaustive enumerations to have a unique minimum-energy structure~\cite{Irback:02,Holzgrafe:11}.
The sequences are labeled S$_N$, where $N$ indicates the number of beads.

A sequence having a unique minimum-energy structure is said to design that structure. The number
of different sequences designing a given structure is called the designability of the structure. Structures
with high designability thus show robustness to mutation.

For a given $N\le30$, the selected sequence S$_{N}$ is one of those that fold to
the structure with highest designability for that $N$.

In addition, we included in our study two sequences with $N=48$ and $N=64$, respectively, for
which low-energy structures have been explored by various classical methods.

The sequences S$_N$ can be found in Table~\ref{tab:seq}, along with their minimum energies, $E_{\min}$.
Some of the corresponding structures are shown in Fig.~\ref{fig:appendix_structures}.

%%%%%%%%%%%%%%%%%%%%%%%%%%%%%%%%%%%%%%%%%%%%%%%%%%%%%%%%%%%%%%%%%%%%%%%%%%%%

\section{Results\label{sec:results}}

Using the spin representation of Sec.~\ref{sec:qubo}, we wish to find minimum-energy structures
of given HP sequences by minimizing the total energy $E=\EHP+\sum_i\lambda_i E_i$ (Eq.~\ref{eq:E})
on a quantum annealer. As a first step toward this goal, we investigate the power of the QUBO
approach under classical SA, and how it depends on the Lagrange parameters $\lambda_i$. We next do the same
using the hybrid quantum-classical solver. Finally, we compare the results by using the QPU annealer only.
We find that the hybrid quantum-classical solver outperforms all the other approaches listed above for our
application, in terms of ease of achieving 100\% hit rate and consumed computer time. 

\subsection{Simulated annealing with QUBO encoding\label{sec:sa_results}}

In our binary model, the $\EHP$ energy can become
substantially lower than it is in any proper chain conformations. For this QUBO approach to work, it is
therefore essential that the $\lambda_i$ parameters that force the solutions to be ``legal'' are sufficiently large. On the other hand, by choosing
large $\lambda_i$ values, one risks making the energy landscape rugged, and therefore the
dynamics potentially slow. Hence, the $\lambda_i$ parameters should be neither too large
nor too small.

To gain insight into the behavior of the binary model and its dependence on the $\lambda_i$ parameters,
we conducted a set of classical Monte Carlo-based SA runs (Sec.~\ref{sec:sa_methods}), using the
HP sequences S$_{18}$--S$_{30}$ in Table~\ref{tab:seq}. The runs had a fixed length and were deemed successful
if the final state corresponded to the known minimum-energy structure of the given HP sequence. As expected, in order to have
an acceptable hit rate, it was necessary to choose the $\lambda_i$ parameters with some care.
Nevertheless, without excessive fine-tuning, it was possible to find a single set of parameters,
$\lv=(2.1,2.4,3.0)$, that gave a hit rate  $\gtrsim$0.1 for all the sequences S$_{18}$--S$_{30}$
(see below). We refrained from attempting any further optimization of the parameters,
as the optimal values need not be the same on a quantum annealer. The optimal parameters
would, of course, also depend on both HP sequence and grid size.

\begin{figure}
\centering
   \includegraphics[width=8cm]{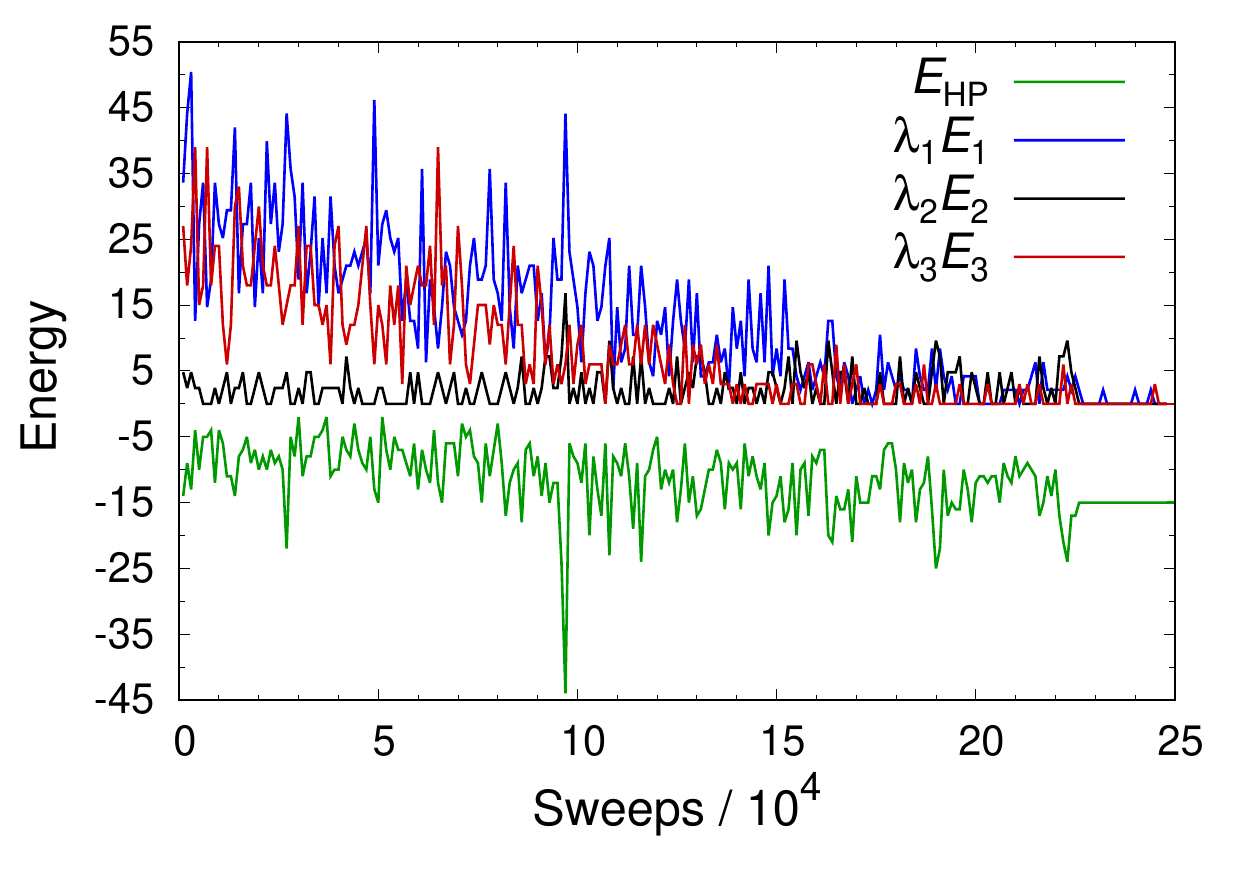}
\caption{Run-time evolution of the HP energy $\EHP$ and the constraint energies
$E_1$, $E_2$ and $E_3$ in a QUBO SA folding simulation for the 30-bead sequence S$_{30}$
(Table~\ref{tab:seq}) on $10^2$ grid, with $\lv=(2.1,2.4,3.0)$.
\label{fig:rt}}
\end{figure}

Figure~\ref{fig:rt} shows the run-time evolution of the four different energy terms in one of
1000 QUBO SA runs for the sequence S$_{30}$.
At the end of the run, all the three constraint energies
$E_i$ vanish, while $\EHP$ takes its known minimum value for an
HP chain with this sequence  ($E_{\min}=-15$). Hence, the final spin configuration
corresponds to the S$_{30}$ ground state in the HP model.
The hit rate, defined as the fraction of runs ending in the ground state,
was $0.226\pm0.013$. The remaining runs ended in spin configurations that
either did not correspond to a proper chain, or corresponded to a structure
with $\EHP>E_{\min}$. In the beginning of the runs, the spin system undergoes a rapid relaxation,
which brings the energies from initial values $\EHP\sim-10^3$ and
$\lambda_1E_1+\lambda_2E_2+\lambda_3E_3\sim10^5$ to
the plotted range before the first measurement is taken (after $10^3$ sweeps).
We note that among the three constraint energies, $E_1$ and $E_3$
tend to relax much more slowly than $E_2$, as is the case in Fig.~\ref{fig:rt}.
Note also that the HP energy takes values $\EHP<E_{\min}$ many times during the course
of the run. Such values can occur only when at least one constraint is broken.

\begin{figure}
\centering
   \includegraphics[width=8cm]{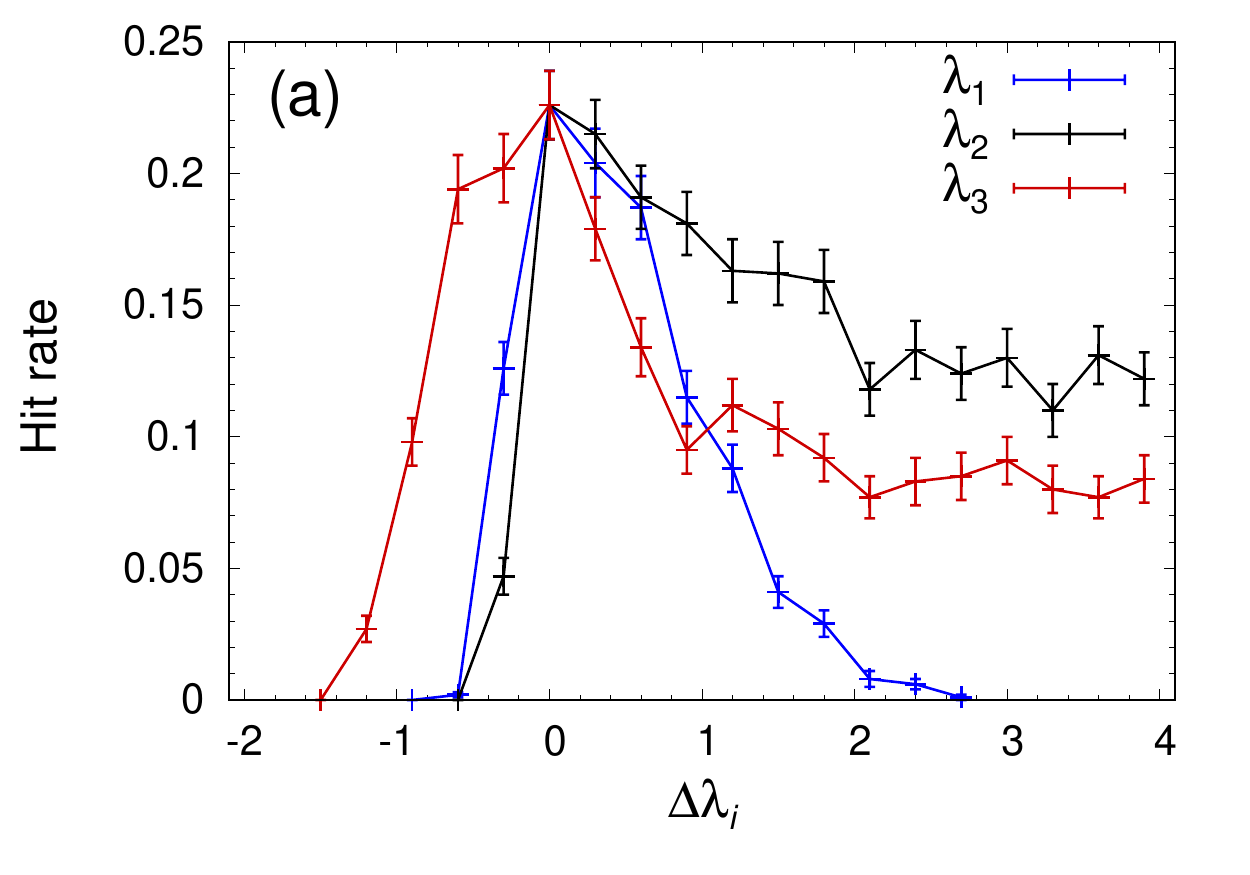}
   \includegraphics[width=8cm]{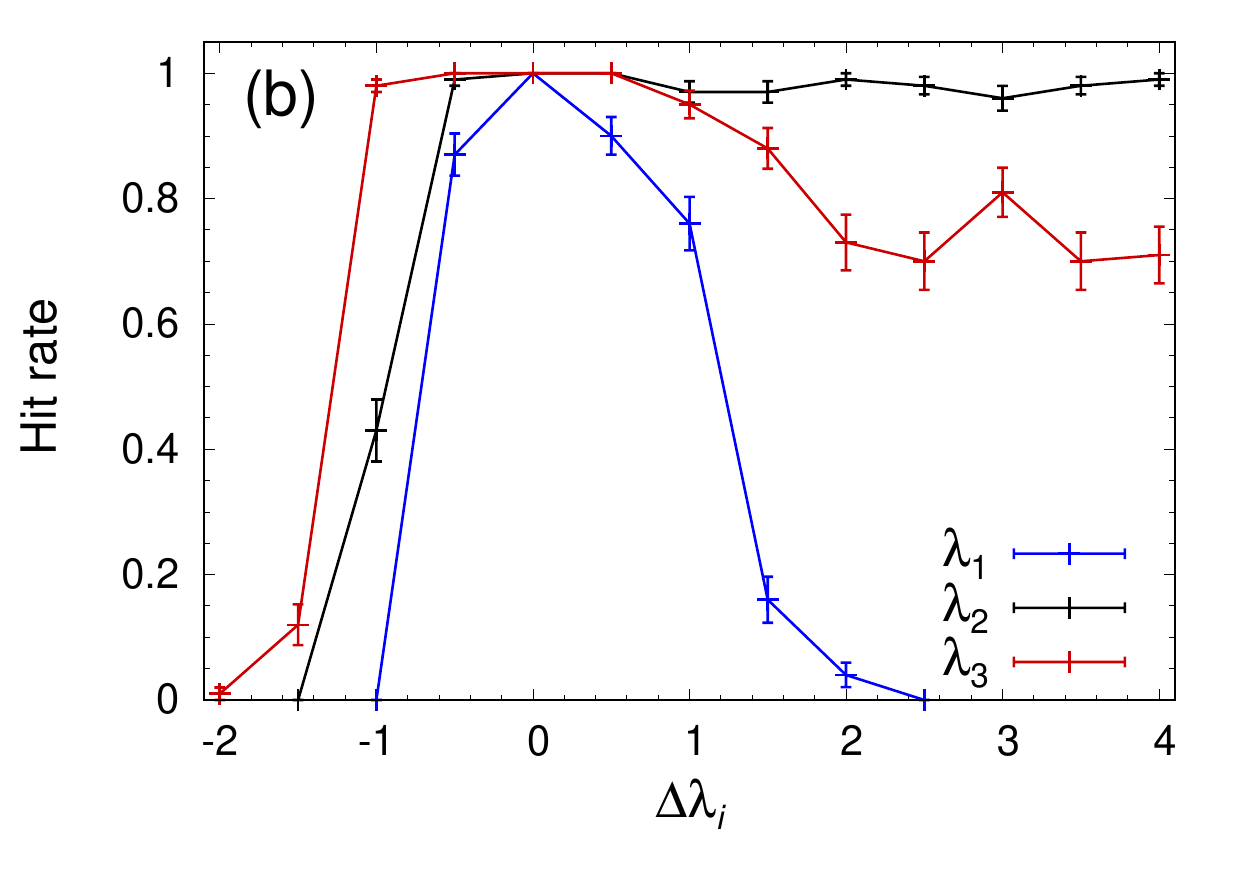}
\caption{Parameter dependence of the fraction of correct solutions (hit rate) in the vicinity of a reference point
$\lv^\ast$, when using QUBO SA and hybrid quantum-classical computation to search for the
ground state of the S$_{30}$ sequence (Table~\ref{tab:seq}) on a $10^2$ grid.
The hit rate is plotted against $\Delta \lambda_i=\lambda_i-\lambda_i^\ast$,
keeping $\lambda_j=\lambda_j^\ast$  for $j\ne i$. Lines are drawn to guide the eye.
(a) QUBO SA with $\lv=(2.1,2.4,3.0)$.
(b) Hybrid quantum-classical computations with $\lv=(2.0,3.0,3.0).$
Note the difference in scale between the two panels, reflecting the difference in performance as shown in Fig.~\ref{fig:sa-hs}.
\label{fig:sensitivity}}
\end{figure}

To elucidate the $\lv$ dependence of the hit rate, we also performed 
QUBO SA calculations for a set of additional $\lv$ near  $\lv = \lv^\ast=(2.1,2.4,3.0)$
using the sequence S$_{30}$. Figure~\ref{fig:sensitivity}(a) shows the observed 
the hit rates when changing one $\lambda_i$ at a time.
In all three $\lambda_i$, the hit rate stays tiny until a threshold is passed,
followed by a steep increase to the maximum observed hit rate, for $\lambda_i = \lambda_i^\ast$.
When further increasing $\lambda_i$ beyond $\lambda_i^\ast$, the hit rate decays, probably due to
an increasingly rugged energy landscape. This decay leads to an upper limit on the parameter $\lambda_1$,
beyond which the hit rate is impractically small. By contrast, the hit rate stays significant
even for $\lambda_2$ and $\lambda_3$ values much larger than those in Fig.~\ref{fig:sensitivity}(a).
In fact, setting $\lambda_2=100$ or $\lambda_3=100$, we still obtained hit rates of $0.132\pm0.011$ and
$0.074\pm 0.008$, respectively. Hence, overall the parameter sensitivity is low, although
$\lambda_1$ must be chosen with some care.

The fact that the $\lambda_1$ dependence has a different shape than the dependencies
on $\lambda_2$ and $\lambda_3$ can be, at least in part, understood. With the single-spin
updates employed, the system cannot move from one chain-like configuration to another, both
with $E_1=E_2=E_3=0$, without visiting intermediate non-chain configurations with $E_1>0$. By contrast,
$E_2$ and $E_3$ may stay zero during such a move. These observations suggest that the
energy landscape indeed is rugged for large $\lambda_1$, but not necessarily so for
large $\lambda_2$ or $\lambda_3$.

\begin{figure}
\centering
   \includegraphics[width=8cm]{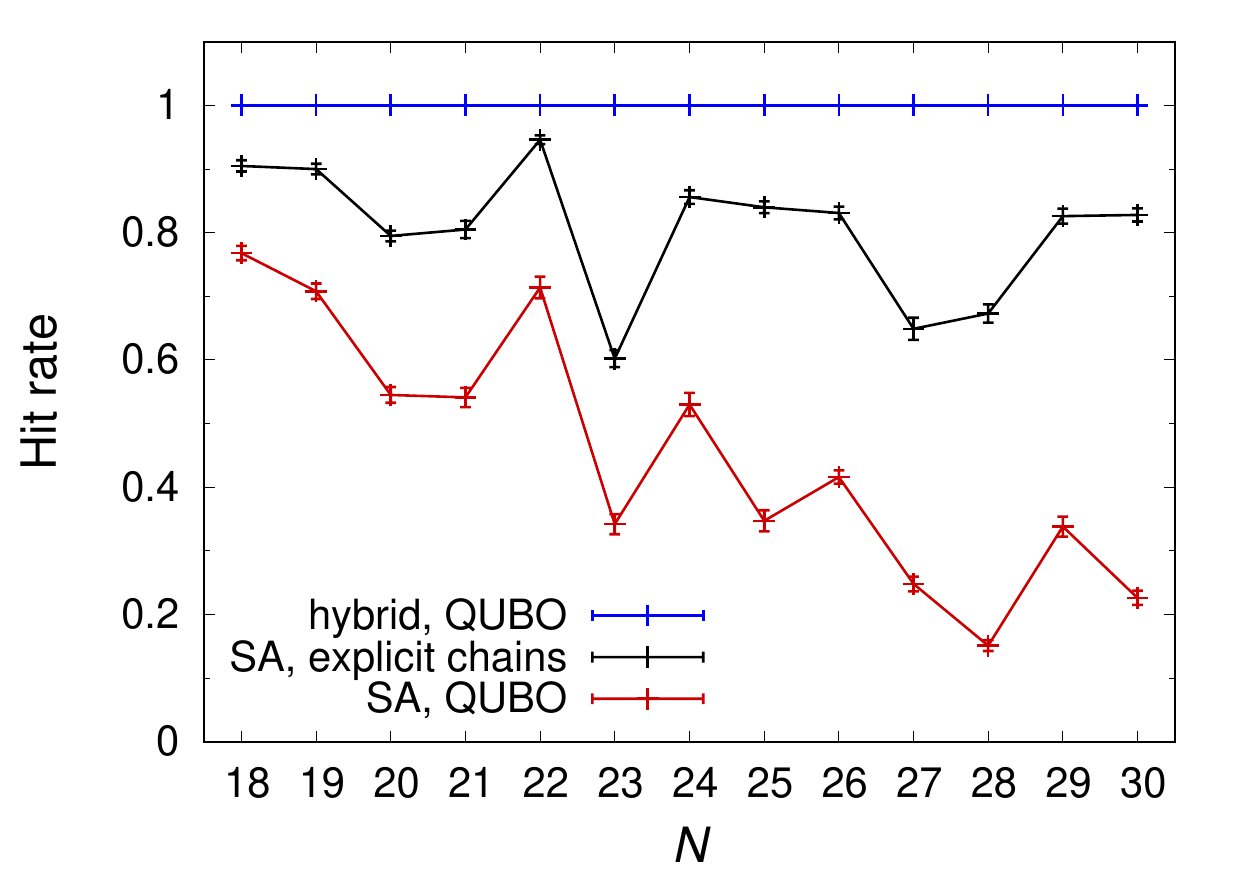}
\caption{Fraction of correct solutions (hit rate) when searching for the ground state of
HP chains with $18\le N\le30$ beads by using the D-Wave hybrid solver with QUBO encoding (blue),
SA with explicit chains (black) and SA with QUBO encoding (red). The HP sequences studied can
be found in Table~\ref{tab:seq}. The parameters $\lv$
were set to $(2.1,2.4,3.0)$ in the QUBO SA runs, and to $(2.0,3.0,3.0)$ when using the
hybrid solver. All QUBO-based results were obtained using a $10^2$ grid.
The hybrid computations used the default run time for the solver (4\,s for $N=30$).
\label{fig:sa-hs}}
\end{figure}

To explore how the hit rate of the QUBO SA approach depends on chain length,
we conducted calculations for all the HP sequences S$_{18}$--S$_{30}$ in Table~\ref{tab:seq},
using $\lv = \lv^\ast$. As expected, the measured hit rates show a decreasing
trend with increasing $N$ (Fig.~\ref{fig:sa-hs}). However, the decrease is not monotonous, indicating that
the hit rate is sequence-dependent and not a simple function of $N$.

For comparison,
we also carried out a set of direct SA minimizations of $\EHP$ based
on conventional explicit-chain Monte Carlo methods (Fig.~\ref{fig:sa-hs}). Despite being faster, the hit
rate is higher in these runs than it is with QUBO-based SA. However, the difference in
hit rate is modest given that state space for explicit chains is comparatively tiny. Note
the similarities in shape between the hit rates obtained from these two unrelated
sets of SA calculations. These similarities suggest that some target structures are
relatively easy or difficult to find, independent of the method employed.

\subsection{Hybrid quantum-classical computations\label{sec:hybrid_results}}

A promising alternative to pure QA is provided by hybrid quantum-classical methods, by which
larger systems can be studied. To assess the power of this approach, we conducted hybrid computations
for all the HP sequences studied in Sec.~\ref{sec:sa_results}, S$_{18}$--S$_{30}$ (Table~\ref{tab:seq}),
using the default run time for the hybrid solver.
We additionally included a two longer sequences~\cite{Unger:93}, which have been extensively
used as testbeds for various (classical) methods. For these sequences, the dependence of the hit rate
on run time was explored.

As in the SA case, with the hybrid solver, a rough search was sufficient in order to find a single
$\lv$, $\lv^\ast=(2.0,3.0,3.0)$, for which all the sequences S$_{18}$--S$_{30}$ could be correctly
folded on a $10^2$ grid. Figure~\ref{fig:sensitivity}(b) shows the parameter dependence of the hit rate
near $\lv^\ast$ when using the hybrid solver. Compared to QUBO SA (Fig.~\ref{fig:sensitivity}(a)),
the measured hit rates are markedly higher with the hybrid solver (Fig.~\ref{fig:sa-hs}).
At the same time, the $\lambda_i$
dependencies share a similar shape in both cases. In particular, in both cases, the hit rate is more
sensitive to changes in $\lambda_1$ than to changes in $\lambda_2$ or $\lambda_3$. As in the SA case,
$\lambda_2$ or $\lambda_3$ can be chosen far above the plotted range in
Fig.~\ref{fig:sensitivity}, without any major loss in hit rate. In fact, when setting $\lambda_2=100$ or
$\lambda_3=100$, we obtained hit rates of $0.980\pm0.014$ and  $0.54\pm 0.05$, respectively.   Overall, the
parameter sensitivity is lower with the hybrid solver than with QUBO SA.

When comparing hit rates from our hybrid and QUBO SA runs for the sequences S$_{18}$--S$_{30}$,
we find that it is consistently highest in the hybrid case (Fig.~\ref{fig:sa-hs}). In fact, the hit rate
is one across this entire set of sequences for the hybrid solver, despite using the shortest
run time that can be set for the solver.

It is important to note that when folding the sequences S$_{18}$--S$_{30}$, the hybrid
solver did not always make use of the QPU. The fraction of runs that used the QPU increased
with $N$ and was above one half for $N>21$. Still, the precise contribution of the QPU to the final results
is hard to judge since the details of the hybrid solver are not publicly available information. Nevertheless,
the ease with which these sequences could be folded motivated us to
also test the hybrid solver on two significantly longer sequences, namely S$_{48}$ and S$_{64}$
(Table~\ref{tab:seq}) with 48 and 64 beads, respectively.

For these two sequences exact results are not available, but both belong to a set of HP
sequences that have been widely used to test novel (classical) algorithms~\cite{Unger:93}.
The lowest known energies are $\EHP=-23$ for S$_{48}$~\cite{Liang:01}
and $\EHP=-42$ for S$_{64}$~\cite{Bastolla:98}. 
In order to obtain good results, the $\lambda_i$ parameters had be to adjusted
for these larger chains to  $\lv=(2.0,3.5,3.0)$ for S$_{48}$ 
and $\lv=(3.0,4.0,4.0)$ for S$_{64}$. In addition, 
using the default run time for the hybrid solver (6\,s for S$_{48}$ and 8\,s for S$_{64}$), as was done for S$_{18}$--S$_{30}$,
turned out to give unsatisfactorily low hit rates for S$_{48}$ and S$_{64}$. Therefore, 
we investigated the run time dependence of the hit rate. To this end, 
we carried out computations with several different run times for both S$_{48}$ and S$_{64}$,
using both $10^2$ and $15^2$ grids. 

\begin{figure}
\centering
\makebox[0.45\columnwidth]{
     \includegraphics[width=8cm,valign=c]{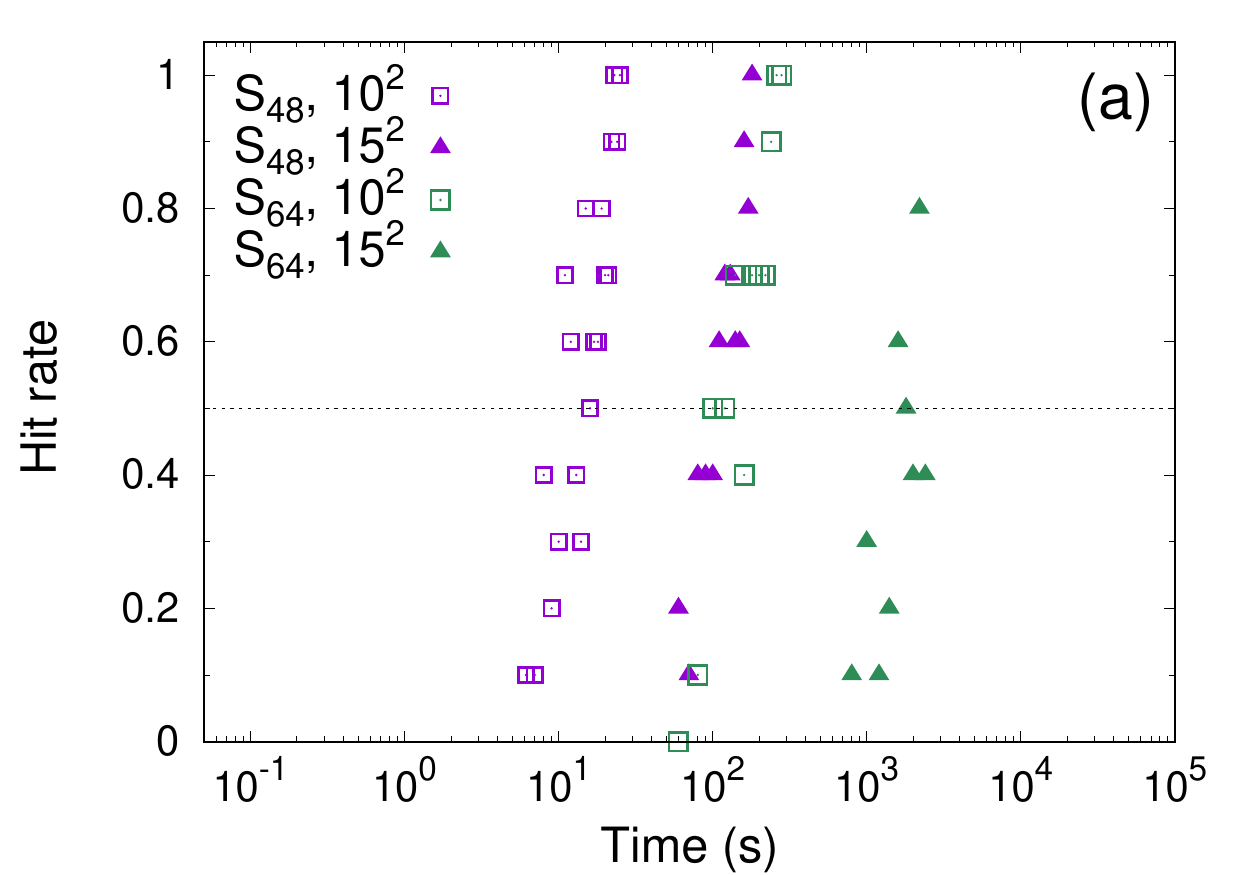}
}
\makebox[0.45\columnwidth]{
     \includegraphics[width=6cm,valign=c]{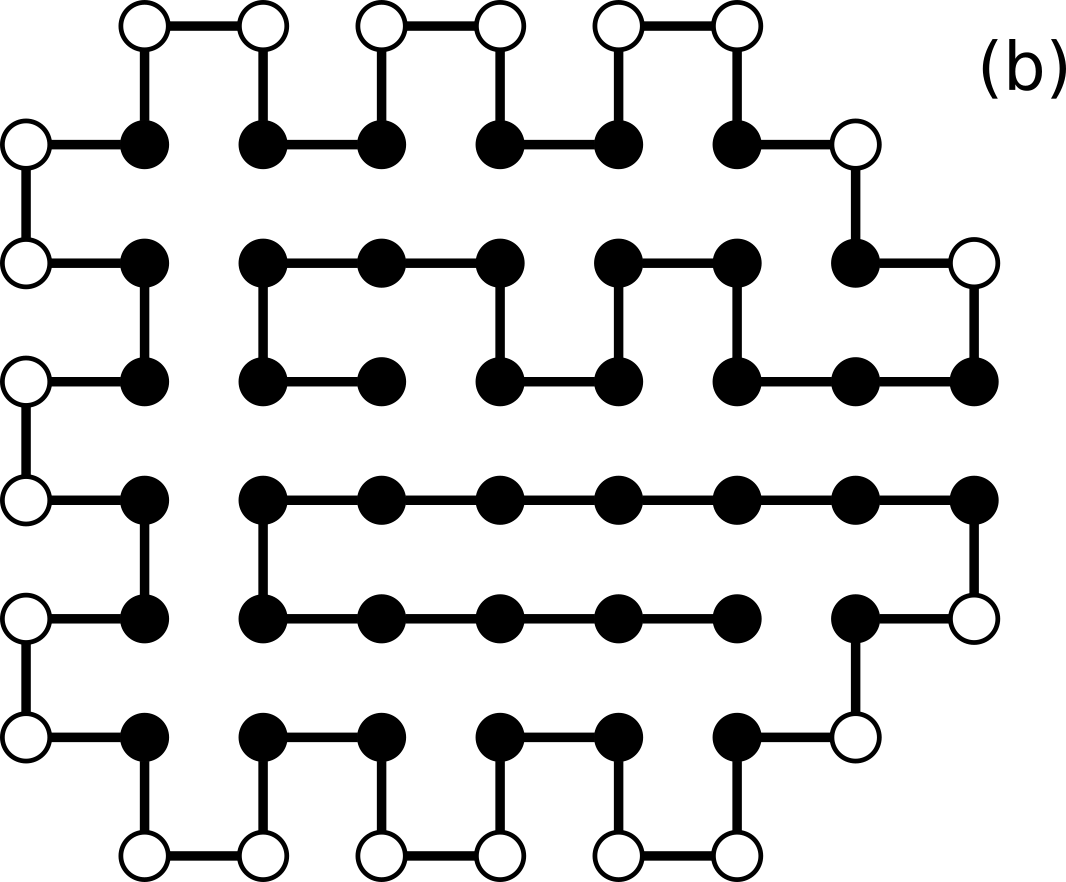}
}
\caption{Hybrid quantum-classical computations for the sequences S$_{48}$ and S$_{64}$ (Table~\ref{tab:seq}), 
using $\lv=(2.0,3.5,3.0)$ for S$_{48}$ and $\lv=(3.0,4.0,4.0)$ for S$_{64}$. 
(a) Fraction of runs that found the lowest known energy, $E_{\min}$, in hybrid computations 
on $10^2$ and $15^2$ grids for the sequences S$_{48}$ ($E_{\min}=-23$~\cite{Liang:01}) and 
S$_{64}$ ($E_{\min}=-42$~\cite{Bastolla:98}),
plotted against run time for the hybrid solver (logarithmic scale). The
horizontal line indicates a hit rate of 0.5.
(b) Example of an S$_{64}$ structure with the lowest known energy ($E_{\min}=-42$) that was 
 obtained in the hybrid computations. Filled and open symbols indicate H and P beads, respectively.
 \label{fig:64chain-hs}}
\end{figure}

Figure~\ref{fig:64chain-hs}(a)
summarizes the run time dependence of the hit rate
as observed in these runs, for both sequences and grids.
For a given sequence and 
grid, a steep increase in hit rate can be seen once the run time passes a threshold. 
While a hit rate of unity is reachable for 
the systems in Fig.~\ref{fig:64chain-hs}(a), a natural measure of performance is 
provided by the 
run time required to obtain a hit rate of 0.5, denoted by $t_{1/2}$. 
For a given grid, the value of $t_{1/2}$ is 
roughly 10 times larger for S$_{64}$ than for S$_{48}$. An increase in $t_{1/2}$ by 
roughly a factor 10 is also observed when increasing the grid size from $10^2$ to
$15^2$ for a given sequence. It would be interesting to further explore the dependence
of $t_{1/2}$ on chain length and grid size. However, a systematic study of this problem
is beyond the scope of the present paper, especially since such a study should also 
address the sequence dependence of $t_{1/2}$. For the sequences
S$_{18}$--S$_{30}$ studied earlier, the default run time is apparently above the threshold, 
as the hit rate is 1.0. 

An example of an S$_{64}$ structure with the lowest known energy that was obtained 
in our hybrid computations can be found in Fig~\ref{fig:64chain-hs}(b). 

\subsection{Pure QPU computations\label{sec:QPU_results}}

The QUBO problem that we wish to solve for finding minimum-energy HP structures
contains $\approx$$NL^2/2$ logical qubits. Moreover, the system is almost fully
connected, implying that its embedding into the QPU topology requires a significant amount of
additional qubits. Therefore, pure QPU computation is effectively limited to relatively short HP chains.

To explore how the performance of the pure QPU approach depends on system size, we conducted
computations for the six sequences S$_4$, S$_6$, S$_7$, S$_8$, S$_9$ and S$_{10}$ (Table~\ref{tab:seq}), 
with each sequence run on several different grids. 
The grids used, which include non-square rectangular ones, can be found in Table~\ref{tab:grids}, 
along with the numbers of logical and physical qubits involved. The number of physical qubits grows to a
good approximation quadratically with the number of logical qubits, 
as illustrated in Fig.~\ref{fig:appendix_physical_and_logical}.

\begin{figure}
\centering
\makebox[0.45\columnwidth]{
     \includegraphics[width=8cm,valign=c]{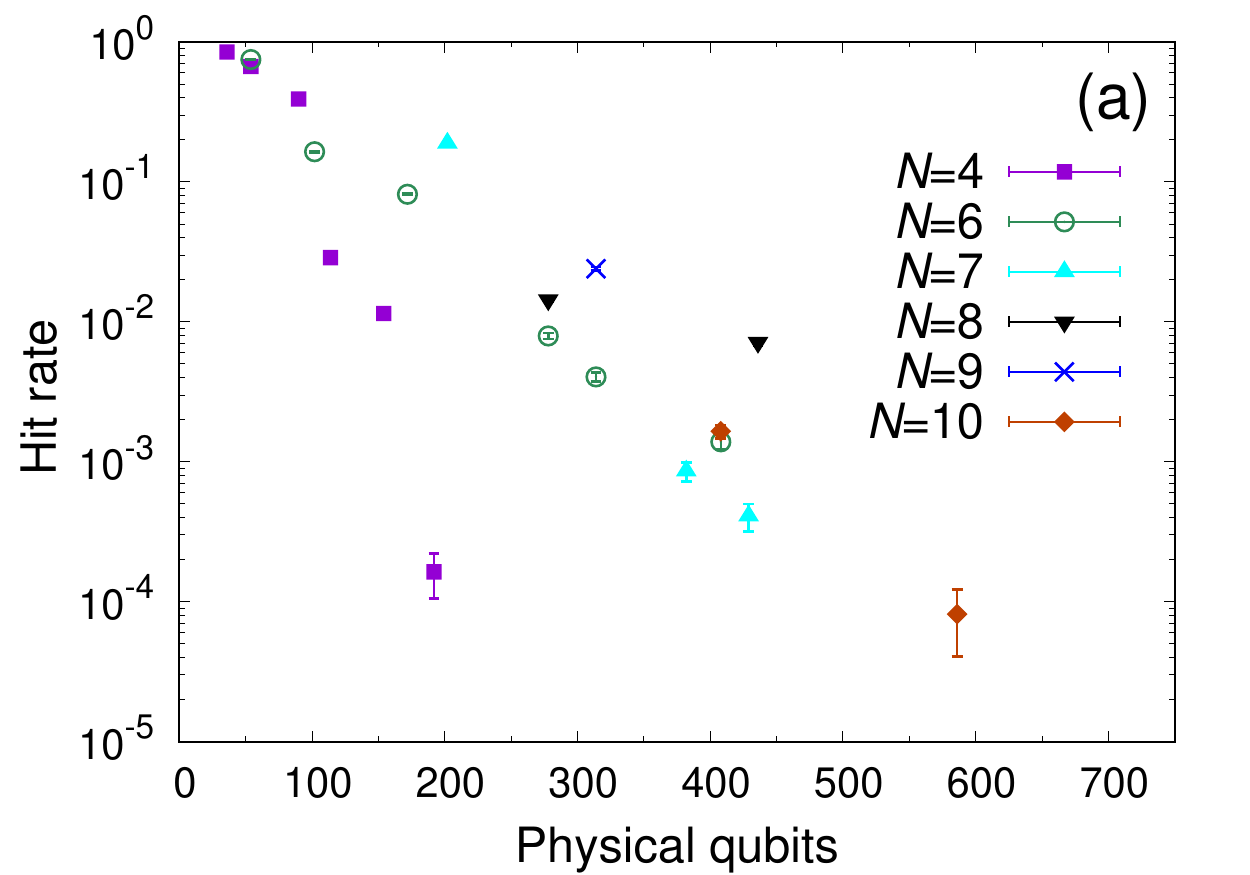}
}
\makebox[0.45\columnwidth]{
     \includegraphics[width=4.0cm,valign=c]{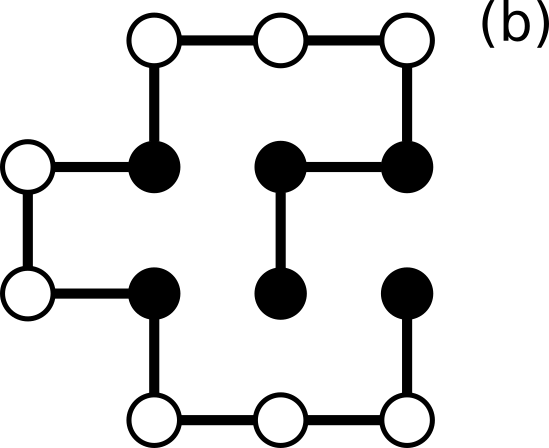}
}
\caption{Pure QPU computations. For every choice of sequence and grid size studied, we conducted 100 runs with 490 annealing cycles each.
The sequences can be found in Table~\ref{tab:seq}.
(a) The hit rate on a logarithmic scale
against the number of physical qubits used for the sequences
S$_4$, S$_6$, S$_7$, S$_8$, S$_9$ and S$_{10}$ for various grid sizes,
using $\lv=(1.0,2.0,1.5)$. The grids used can be found in Table~\ref{tab:grids}.
Statistical errors are in many cases comparable with or smaller than the symbol sizes.
(b) The minimum-energy structure for the sequence S$_{14}$, which was successfully recovered on a $4^2$ grid, using $\lv=(2.0,7.0,4.0)$.
Filled and open symbols indicate H and P beads, respectively.
\label{fig:pureQPU}}
\end{figure}

Figure~\ref{fig:pureQPU}(a) shows the fraction of all annealing cycles that recovered
the known minimum-energy structure, for these systems, plotted against the number of
physical qubits employed. The parameters $\lambda_i$ and the annealing time were the same for all systems,
whereas the chain strength was chosen individually for each system, for best performance (among the values
1.0, 1.5,\ldots, 4.5, 5.0). Albeit with some scatter, the hit rate shows a roughly exponential decay
with the number of physical qubits used. In part, this deterioration of 
performance with increasing system size may stem from integrated Hamiltonian control errors, which will reduce
the probability of finding the ground state of the intended Hamiltonian~\cite{Pearson:19}. Remedies 
to this problem are being explored~\cite{Pearson:19}.

In Fig.~\ref{fig:pureQPU}(a), the hit rate falls off most rapidly for the shortest chain studied, with $N=4$.  
This behavior is likely an artifact due to suboptimal $\lambda_i$ parameters, which, for simplicity, are
kept the same for all systems. In fact, for the largest $N=4$ system with 192 physical qubits, 
a significantly higher hit rate ($\approx$0.02) was observed when changing 
$\lambda_1$ from its value 1.0 in Fig.~\ref{fig:pureQPU}(a) to 1.5.       

The longest sequence that was successfully folded in our pure QPU computations
was S$_{14}$ with 14 beads (Table~\ref{tab:seq}). This sequence, whose minimum-energy structure can
be seen in Fig.~\ref{fig:pureQPU}(b), was studied using a
$4^2$ grid, which required 112 logical and 1214 physical qubits. The chain strength was set to 7.5.
To our knowledge, this is the largest protein succesfully folded using a quantum computer.
However, the ground state was only recovered in one of a total of $100\times 490$ annealing cycles. This
number of cycles is larger than the number of distinct conformations available to a chain with 14 beads
on a $4^2$ grid, which is 416. On the other hand, it 
is tiny compared to the $2^{112}\approx 5.2\times 10^{33}$ states of the binary system, the vast majority of
which do not correspond to proper chain configurations. Finally, we note that our S$_{14}$ system is similar in size
to the largest system solved in a recent benchmarking study of the D-Wave Advantage machine using exact 
cover problems, which contained 120 logical qubits~\cite{Willsch:22}. 

It is possible that the pure QPU results can improved by further tuning of the simulation
parameters. However, at present, we conclude that pure QPU computation cannot match
classical SA or the hybrid quantum-classical approach (Secs.~\ref{sec:sa_results}, \ref{sec:hybrid_results}).

\section{Summary and Outlook}

We have developed a novel mapping of the two-dimensional HP lattice protein model onto a quantum annealer, which is successfully
explored on the D-Wave Advantage system. This simplified model for protein structures is known to represent a difficult
optimization problem when determining structure using classical methods. As benchmarks for success, we used HP
chains with sizes ranging from $N=4$ to $N=30$, for which the exact solutions are known from exhaustive enumerations.
For larger problems, with $N=48$ and $N=64$, we compared with the best known solutions obtained by classical means. The
approach allows us to explore the largest chains studied so far on a quantum annealer, and consistently provides high percentages of
correct solution on multiple runs. However, the success of our approach relies upon using the hybrid variant provided by D-Wave. The
performance of pure QA is less impressive with a drastic decrease in success rate as the system size is increased. These
calculations were therefore limited to chains with at most $N=14$.

Our approach differs from previous attempts based upon growth algorithms, as it maps the problem onto a lattice spin system
where the spins, or qubits, are present throughout the lattice.
In comparison to earlier work, this representation greatly facilitates the handling of interactions, including self-avoidance.
%Furthermore, except for very short chains, this representation is more resource-efficient, as the amount
%of logical qubits required exhibits a polynomial rather than exponential scaling with chain length.
In particular, with this encoding, the energy function can be written in a quadratic form
without adding any auxiliary spins.  

The encoding requires penalty terms for the global energy minimum of the spin system to correspond to a proper chain.
The method is robust to changes in the strengths of the penalty terms, which require only a modest amount
of tuning.

For convenience, we have focused on the HP model with its minimal two-letter alphabet. To extend the approach
to models with larger alphabets, such as the 20-letter Miyazawa-Jernigan model~\cite{Miyazawa:96}, is straightforward,
as it amounts to simply changing the interaction parameters in Eq.~\ref{eq:EHP}.
Moreover, although the checkerboard division into even and odd sites may have to be modified or abandoned,
the method can be applied to an arbitrary graph. In particular, it can be directly applied to
three-dimensional grids, including the tetrahedral one used in Ref.~\cite{Robert:21}.

Finally, we note that a similar approach could be applicable to gate-based quantum computers. This could potentially take
the form of a quantum variational algorithm or a quantum search algorithm.

\clearpage

\appendix*

\section{}

\setcounter{table}{0}
\renewcommand{\thetable}{A\arabic{table}}
\begin{table}[H]
  \centering
  \caption{The HP sequences studied. The sequences are labeled S$_N$, where
  $N$ indicates the number of beads. All S$_N$ with $N\le 30$ have a known, unique
  minimum-energy structure~\cite{Irback:02,Holzgrafe:11}. The minimum energy is denoted
  by $E_{\min}$. For all $N\le30$, the sequence S$_{N}$ is chosen
  among those having the most highly designable structure for this $N$ as its unique minimum-energy
  structure. For the additional and
  longer sequences S$_{48}$ and S$_{64}$, the ground states are unknown. Here, the $E_{\min}$ values,
  marked with an asterisk, are the currently lowest known energies, found with classical
  methods~\cite{Liang:01,Bastolla:98}. Low-energy structures for the different sequences can
  be found in Figs.~\ref{fig:64chain-hs}(b) (S$_{64}$), \ref{fig:pureQPU}(b) (S$_{14}$), and \ref{fig:appendix_structures} (all other S$_N$).
  \label{tab:seq}}
  \vspace{6pt}
  \begin{tabular}{llcc}
    \hline
    Name & Sequence & $E_{\min}$ \\
    \hline
    \vspace{-4pt}
    S$_4$     & HPPH									& $-1$\\
    \vspace{-4pt}
    S$_6$     & HPPHPH								& $-2$\\
    \vspace{-4pt}
    S$_7$     & PHPPHPH								& $-2$\\
    \vspace{-4pt}
    S$_8$     & HPHPHPPH							& $-3$\\
    \vspace{-4pt}
    S$_9$	   & HHPPHPPHP							& $-3$\\
    \vspace{-4pt}
    S$_{10}$ & HPPHPPHPPH              					& $-4$ \\
    \vspace{-4pt}
    S$_{14}$ & HHHPPPHPPHPPPH        					& $-5$ \\
    \vspace{-4pt}
    S$_{18}$ & HHHPPHPPHPHPPHPHPH   				& $-9$ \\
    \vspace{-4pt}
    S$_{19}$ & PHPHPHPPHPHPPHPPHHH 				& $-9$ \\
    \vspace{-4pt}
    S$_{20}$ & HPHPHPPHPHPPHPPPPHHH 				& $-9$ \\
    \vspace{-4pt}
    S$_{21}$ & PHHPPHPHPPHPHPPHPPHHH 			& $-10$ \\
    \vspace{-4pt}
    S$_{22}$ & HPPHPPHPHPPHPHPPHPPHHH 			& $-11$ \\
    \vspace{-4pt}
    S$_{23}$ & PPHHHHPPHPPHPHPPHPHPPHP 			& $-10$ \\
    \vspace{-4pt}
    S$_{24}$ & HPPPPHPPHPHPPHPHPPHPPHHH 			& $-11$ \\
    \vspace{-4pt}
    S$_{25}$ & PHPHPHPHPPHPHPHPPHPPHHHHH 		& $-13$ \\
    \vspace{-4pt}
    S$_{26}$ & HHHHPPHHPPHPHPPHPHPPHHPPHH 		& $-14$ \\
    \vspace{-4pt}
    S$_{27}$ & PHPHPHPHPPHPHPHPPHPPPPHHHHH 		& $-13$ \\
    \vspace{-4pt}
    S$_{28}$ & PPHHHPPHPPHPHPHPPHPHPPHPPHHH 	& $-13$ \\
    \vspace{-4pt}
    S$_{29}$ & PHPHPHPPHHPPHPHPPHPPHHHHPPHHH 	& $-15$ \\
    \vspace{-4pt}
    S$_{30}$ & PPHHHHPPHPPHPHPPHHPPHPHPHPPHHH 	& $-15$ \\
    \vspace{-4pt}
    S$_{48}$ & PPHPPHHPPHHPPPPPHHHHHHHHHHPPPPPPHHPPHHPPHPPHHHHH & $-23^\ast$ \\
    \vspace{-6pt}
    S$_{64}$ & HHHHHHHHHHHHPHPHPPHHPPHHPPHPPHHPPHHPPHPPHHPPHHPP-\\
    		   & HPHPHHHHHHHHHHHHH & $-42^\ast$\\
  	\hline
  \end{tabular}
\end{table}

\begin{table}[H]
  \centering
  \caption{The grids used when obtaining the data in Fig.~\ref{fig:pureQPU}(a) 
  for the sequences S$_4$, S$_6$, S$_7$, S$_8$, S$_9$ and S$_{10}$ (Table~\ref{tab:seq}), 
  along with the numbers of logical and physical qubits needed.  
  The embeddings into the QPU topology were generated using \texttt{DWaveCliqueSampler}.
  \label{tab:grids}}
  \vspace{6pt}
  \begin{tabular}{l@{\hspace{.5cm}}c@{\hspace{.5cm}}c@{\hspace{.5cm}}c}
    \hline
    Sequence & Grid & Logical qubits & Physical qubits\\
    \hline    
    S$_4$      & $3\times2$ & 12 & 36\\ 
                    & $3\times3$ & 18 & 54\\
                    & $4\times3$  & 24 & 90\\
                    & $5\times3$  & 30 &114\\
                    & $4\times4$ & 32 & 154\\
                    & $5\times4$ & 40 & 192\\
    \hline
    S$_6$     & $3\times2$ &18 & 54\\
                    & $3\times3$ & 27 & 102\\
                    & $4\times3$ & 36 & 172\\
                    & $4\times4$ & 48 & 278\\
                    & $6\times3$ & 54 & 314\\
     \hline    	
     S$_7$     & $4\times3$ & 42 & 202 \\
                    & $4\times4$ &  56 & 382 \\
                    & $6\times3$ & 63 & 429 \\
    \hline
    S$_8$     & $4\times3$ & 48 & 278 \\ 
                   &  $4\times4$ & 64 & 436 \\
    \hline
    S$_9$     & $4\times3$ & 54 & 314\\
    \hline
    S$_{10}$     & $4\times3$ & 60 & 408 \\
                        & $5\times3$ & 75 & 586\\   
     \hline
  \end{tabular}
\end{table}

\renewcommand{\thefigure}{A\arabic{figure}}
\setcounter{figure}{0}
\begin{figure}[H]
    \centering
    \includegraphics[width=11cm]{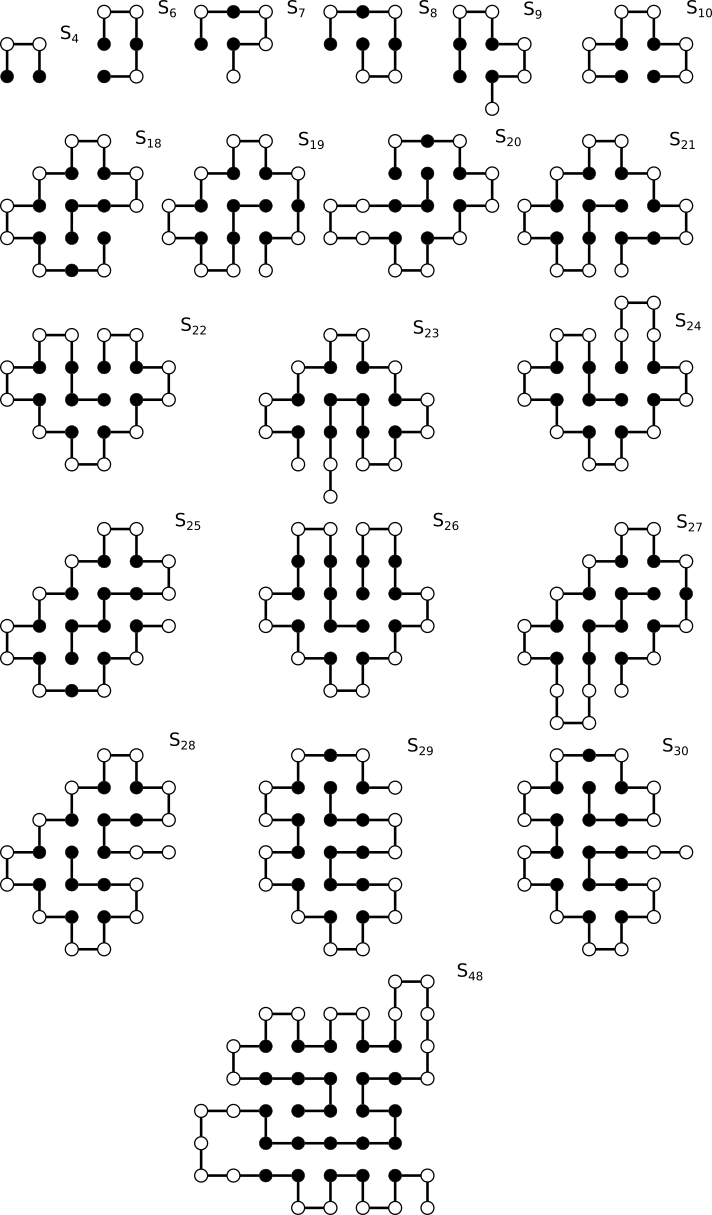}
    \caption{Ground states for all the sequences S$_N$ in Table~\ref{tab:seq} with $N\le 30$~\cite{Irback:02,Holzgrafe:11}
    except S$_{14}$ (whose ground state can be found in Fig.~\ref{fig:pureQPU}(b)). Also shown is an S$_{48}$ structure, which is one of the
    structures with the lowest known energy for this sequence found using the hybrid solver. A low-energy
    structure for the sequence S$_{64}$ can be found in Fig.~\ref{fig:64chain-hs}(b).}
    \label{fig:appendix_structures}
\end{figure}

\begin{figure}[H]
    \centering
    \includegraphics[width=8cm]{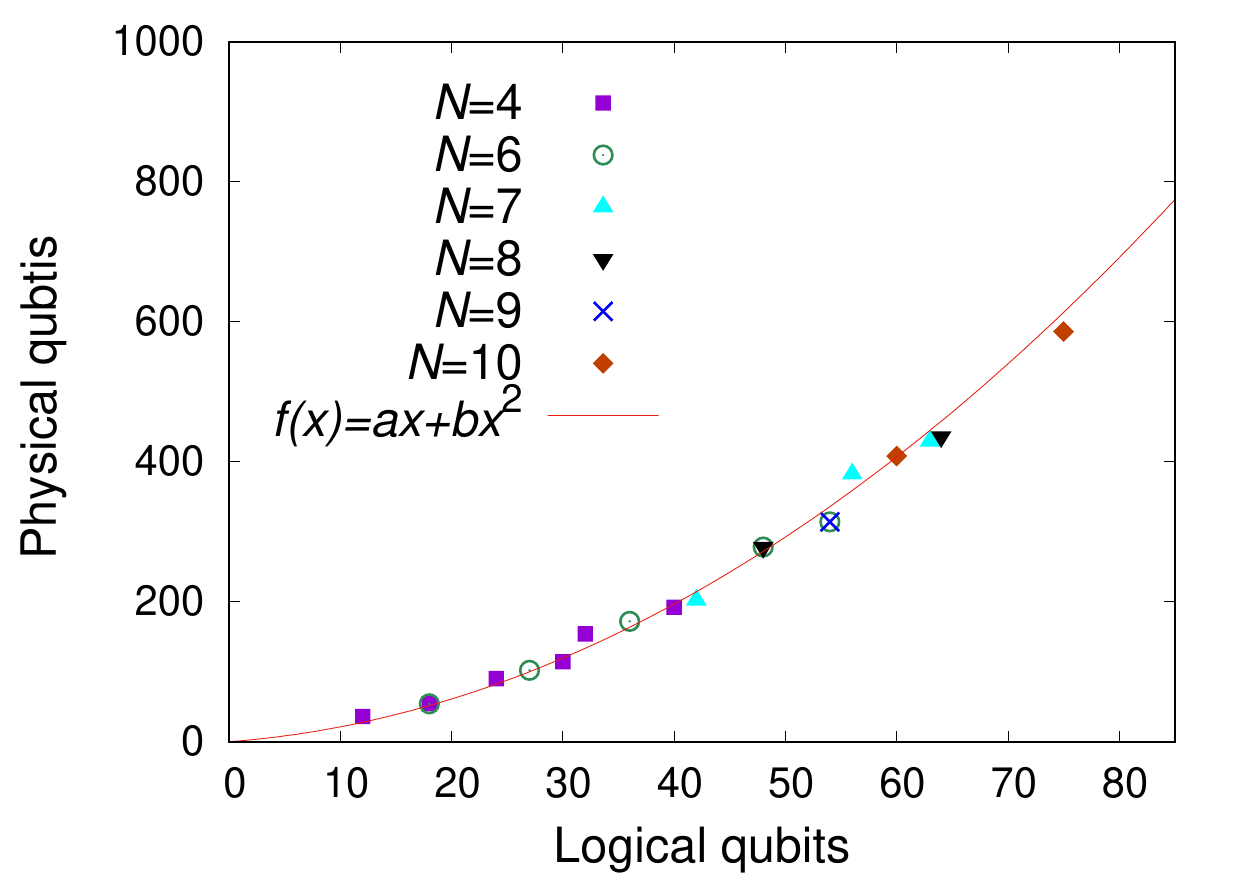}
    \caption{Relation between the numbers of physical and logical qubits in pure QPU calculations 
    for the sequences S$_4$, S$_6$, S$_7$, S$_8$, S$_9$ and S$_{10}$ on different grids 
    (Fig.~\ref{fig:pureQPU}(a), Table~\ref{tab:grids}). 
    The curve represents a fit of the form $f(x)=ax+bx^2$ ($a=1.85$, $b=0.093$). The embeddings into the 
    QPU topology were generated using \texttt{DWaveCliqueSampler}.}
    \label{fig:appendix_physical_and_logical}
\end{figure}

%\section{}

% If you have acknowledgments, this puts in the proper section head.

\begin{acknowledgments}
This work was in part supported by the Swedish
Research Council (Grant no.~621-2018-04976).
We gratefully acknowledge the J\"ulich Supercomputing Centre
(https://www.fz-juelich.de/ias/jsc) for supporting this project by providing computing time
on the D-Wave Advantage\texttrademark{} System JUPSI through the J\"ulich UNified
Infrastructure for Quantum computing (JUNIQ).
We have enjoyed fruitful discussions with theory group members at the Wallenberg Centre
for Quantum Technology at Chalmers University. In particular, we would like to thank Hanna Linn for feedback on the code.
\end{acknowledgments}

% Create the reference section using BibTeX:
%\bibliography{../refs}

\begin{thebibliography}{28}%
\makeatletter
\providecommand \@ifxundefined [1]{%
 \@ifx{#1\undefined}
}%
\providecommand \@ifnum [1]{%
 \ifnum #1\expandafter \@firstoftwo
 \else \expandafter \@secondoftwo
 \fi
}%
\providecommand \@ifx [1]{%
 \ifx #1\expandafter \@firstoftwo
 \else \expandafter \@secondoftwo
 \fi
}%
\providecommand \natexlab [1]{#1}%
\providecommand \enquote  [1]{``#1''}%
\providecommand \bibnamefont  [1]{#1}%
\providecommand \bibfnamefont [1]{#1}%
\providecommand \citenamefont [1]{#1}%
\providecommand \href@noop [0]{\@secondoftwo}%
\providecommand \href [0]{\begingroup \@sanitize@url \@href}%
\providecommand \@href[1]{\@@startlink{#1}\@@href}%
\providecommand \@@href[1]{\endgroup#1\@@endlink}%
\providecommand \@sanitize@url [0]{\catcode `\\12\catcode `\$12\catcode
  `\&12\catcode `\#12\catcode `\^12\catcode `\_12\catcode `\%12\relax}%
\providecommand \@@startlink[1]{}%
\providecommand \@@endlink[0]{}%
\providecommand \url  [0]{\begingroup\@sanitize@url \@url }%
\providecommand \@url [1]{\endgroup\@href {#1}{\urlprefix }}%
\providecommand \urlprefix  [0]{URL }%
\providecommand \Eprint [0]{\href }%
\providecommand \doibase [0]{https://doi.org/}%
\providecommand \selectlanguage [0]{\@gobble}%
\providecommand \bibinfo  [0]{\@secondoftwo}%
\providecommand \bibfield  [0]{\@secondoftwo}%
\providecommand \translation [1]{[#1]}%
\providecommand \BibitemOpen [0]{}%
\providecommand \bibitemStop [0]{}%
\providecommand \bibitemNoStop [0]{.\EOS\space}%
\providecommand \EOS [0]{\spacefactor3000\relax}%
\providecommand \BibitemShut  [1]{\csname bibitem#1\endcsname}%
\let\auto@bib@innerbib\@empty
%</preamble>
\bibitem [{\citenamefont {Johnson}\ \emph {et~al.}(2011)\citenamefont
  {Johnson}, \citenamefont {Amin}, \citenamefont {Gildert}, \citenamefont
  {Lanting}, \citenamefont {Hamze}, \citenamefont {Dickson}, \citenamefont
  {Harris}, \citenamefont {Berkley}, \citenamefont {Johansson}, \citenamefont
  {Bunyk}, \citenamefont {Chapple}, \citenamefont {Enderud}, \citenamefont
  {Hilton}, \citenamefont {Karimi}, \citenamefont {Ladizinsky}, \citenamefont
  {Ladizinsky}, \citenamefont {Oh}, \citenamefont {Perminov}, \citenamefont
  {Rich}, \citenamefont {Thom}, \citenamefont {Tolkacheva}, \citenamefont
  {Truncik}, \citenamefont {Uchaikin}, \citenamefont {Wang}, \citenamefont
  {Wilson},\ and\ \citenamefont {Rose}}]{Johnson:11}%
  \BibitemOpen
  \bibfield  {author} {\bibinfo {author} {\bibfnamefont {M.~W.}\ \bibnamefont
  {Johnson}}, \bibinfo {author} {\bibfnamefont {M.~H.~S.}\ \bibnamefont
  {Amin}}, \bibinfo {author} {\bibfnamefont {S.}~\bibnamefont {Gildert}},
  \bibinfo {author} {\bibfnamefont {T.}~\bibnamefont {Lanting}}, \bibinfo
  {author} {\bibfnamefont {F.}~\bibnamefont {Hamze}}, \bibinfo {author}
  {\bibfnamefont {N.}~\bibnamefont {Dickson}}, \bibinfo {author} {\bibfnamefont
  {R.}~\bibnamefont {Harris}}, \bibinfo {author} {\bibfnamefont {A.~J.}\
  \bibnamefont {Berkley}}, \bibinfo {author} {\bibfnamefont {J.}~\bibnamefont
  {Johansson}}, \bibinfo {author} {\bibfnamefont {P.}~\bibnamefont {Bunyk}},
  \bibinfo {author} {\bibfnamefont {E.~M.}\ \bibnamefont {Chapple}}, \bibinfo
  {author} {\bibfnamefont {C.}~\bibnamefont {Enderud}}, \bibinfo {author}
  {\bibfnamefont {J.~P.}\ \bibnamefont {Hilton}}, \bibinfo {author}
  {\bibfnamefont {K.}~\bibnamefont {Karimi}}, \bibinfo {author} {\bibfnamefont
  {E.}~\bibnamefont {Ladizinsky}}, \bibinfo {author} {\bibfnamefont
  {N.}~\bibnamefont {Ladizinsky}}, \bibinfo {author} {\bibfnamefont
  {T.}~\bibnamefont {Oh}}, \bibinfo {author} {\bibfnamefont {I.}~\bibnamefont
  {Perminov}}, \bibinfo {author} {\bibfnamefont {C.}~\bibnamefont {Rich}},
  \bibinfo {author} {\bibfnamefont {M.~C.}\ \bibnamefont {Thom}}, \bibinfo
  {author} {\bibfnamefont {E.}~\bibnamefont {Tolkacheva}}, \bibinfo {author}
  {\bibfnamefont {C.~J.~S.}\ \bibnamefont {Truncik}}, \bibinfo {author}
  {\bibfnamefont {S.}~\bibnamefont {Uchaikin}}, \bibinfo {author}
  {\bibfnamefont {J.}~\bibnamefont {Wang}}, \bibinfo {author} {\bibfnamefont
  {B.}~\bibnamefont {Wilson}},\ and\ \bibinfo {author} {\bibfnamefont
  {G.}~\bibnamefont {Rose}},\ }\bibfield  {title} {\bibinfo {title} {Quantum
  annealing with manufactured spins},\ }\href@noop {} {\bibfield  {journal}
  {\bibinfo  {journal} {Nature}\ }\textbf {\bibinfo {volume} {473}},\ \bibinfo
  {pages} {194} (\bibinfo {year} {2011})}\BibitemShut {NoStop}%
\bibitem [{\citenamefont {Arute}\ \emph {et~al.}(2019)\citenamefont {Arute},
  \citenamefont {Arya}, \citenamefont {Babbush}, \citenamefont {Bacon},
  \citenamefont {Bardin}, \citenamefont {Barends}, \citenamefont {Biswas},
  \citenamefont {Boixo}, \citenamefont {Brandao}, \citenamefont {Buell},
  \citenamefont {Burkett}, \citenamefont {Chen}, \citenamefont {Chen},
  \citenamefont {Chiaro}, \citenamefont {Collins}, \citenamefont {Courtney},
  \citenamefont {Dunsworth}, \citenamefont {Farhi}, \citenamefont {Foxen},
  \citenamefont {Fowler}, \citenamefont {Gidney}, \citenamefont {Giustina},
  \citenamefont {Graff}, \citenamefont {Guerin}, \citenamefont {Habegger},
  \citenamefont {Harrigan}, \citenamefont {Hartmann}, \citenamefont {Ho},
  \citenamefont {Hoffmann}, \citenamefont {Huang}, \citenamefont {Humble},
  \citenamefont {Isakov}, \citenamefont {Jeffrey}, \citenamefont {Jiang},
  \citenamefont {Kafri}, \citenamefont {Kechedzhi}, \citenamefont {Kelly},
  \citenamefont {Klimov}, \citenamefont {Knysh}, \citenamefont {Korotkov},
  \citenamefont {Kostritsa}, \citenamefont {Landhuis}, \citenamefont
  {Lindmark}, \citenamefont {Lucero}, \citenamefont {Lyakh}, \citenamefont
  {Mandr\`a}, \citenamefont {McClean}, \citenamefont {McEwen}, \citenamefont
  {Megrant}, \citenamefont {Mi}, \citenamefont {Michielsen}, \citenamefont
  {Mohseni}, \citenamefont {Mutus}, \citenamefont {Naaman}, \citenamefont
  {Neeley}, \citenamefont {Neill}, \citenamefont {Niu}, \citenamefont {Ostby},
  \citenamefont {Petukhov}, \citenamefont {Platt}, \citenamefont {Quintana},
  \citenamefont {Rieffel}, \citenamefont {Roushan}, \citenamefont {Rubin},
  \citenamefont {Sank}, \citenamefont {Satzinger}, \citenamefont {Smelyanskiy},
  \citenamefont {Sung}, \citenamefont {Trevithick}, \citenamefont
  {Vainsencher}, \citenamefont {Villalonga}, \citenamefont {White},
  \citenamefont {Yao}, \citenamefont {Yeh}, \citenamefont {Zalcman},
  \citenamefont {Neven},\ and\ \citenamefont {Martinis}}]{Arute:19}%
  \BibitemOpen
  \bibfield  {author} {\bibinfo {author} {\bibfnamefont {F.}~\bibnamefont
  {Arute}}, \bibinfo {author} {\bibfnamefont {K.}~\bibnamefont {Arya}},
  \bibinfo {author} {\bibfnamefont {R.}~\bibnamefont {Babbush}}, \bibinfo
  {author} {\bibfnamefont {D.}~\bibnamefont {Bacon}}, \bibinfo {author}
  {\bibfnamefont {J.~C.}\ \bibnamefont {Bardin}}, \bibinfo {author}
  {\bibfnamefont {R.}~\bibnamefont {Barends}}, \bibinfo {author} {\bibfnamefont
  {R.}~\bibnamefont {Biswas}}, \bibinfo {author} {\bibfnamefont
  {S.}~\bibnamefont {Boixo}}, \bibinfo {author} {\bibfnamefont {F.~G. S.~L.}\
  \bibnamefont {Brandao}}, \bibinfo {author} {\bibfnamefont {D.~A.}\
  \bibnamefont {Buell}}, \bibinfo {author} {\bibfnamefont {B.}~\bibnamefont
  {Burkett}}, \bibinfo {author} {\bibfnamefont {Y.}~\bibnamefont {Chen}},
  \bibinfo {author} {\bibfnamefont {Z.}~\bibnamefont {Chen}}, \bibinfo {author}
  {\bibfnamefont {B.}~\bibnamefont {Chiaro}}, \bibinfo {author} {\bibfnamefont
  {R.}~\bibnamefont {Collins}}, \bibinfo {author} {\bibfnamefont
  {W.}~\bibnamefont {Courtney}}, \bibinfo {author} {\bibfnamefont
  {A.}~\bibnamefont {Dunsworth}}, \bibinfo {author} {\bibfnamefont
  {E.}~\bibnamefont {Farhi}}, \bibinfo {author} {\bibfnamefont
  {B.}~\bibnamefont {Foxen}}, \bibinfo {author} {\bibfnamefont
  {A.}~\bibnamefont {Fowler}}, \bibinfo {author} {\bibfnamefont
  {C.}~\bibnamefont {Gidney}}, \bibinfo {author} {\bibfnamefont
  {M.}~\bibnamefont {Giustina}}, \bibinfo {author} {\bibfnamefont
  {R.}~\bibnamefont {Graff}}, \bibinfo {author} {\bibfnamefont
  {K.}~\bibnamefont {Guerin}}, \bibinfo {author} {\bibfnamefont
  {S.}~\bibnamefont {Habegger}}, \bibinfo {author} {\bibfnamefont {M.~P.}\
  \bibnamefont {Harrigan}}, \bibinfo {author} {\bibfnamefont {M.~J.}\
  \bibnamefont {Hartmann}}, \bibinfo {author} {\bibfnamefont {A.}~\bibnamefont
  {Ho}}, \bibinfo {author} {\bibfnamefont {M.}~\bibnamefont {Hoffmann}},
  \bibinfo {author} {\bibfnamefont {T.}~\bibnamefont {Huang}}, \bibinfo
  {author} {\bibfnamefont {T.~S.}\ \bibnamefont {Humble}}, \bibinfo {author}
  {\bibfnamefont {S.~V.}\ \bibnamefont {Isakov}}, \bibinfo {author}
  {\bibfnamefont {E.}~\bibnamefont {Jeffrey}}, \bibinfo {author} {\bibfnamefont
  {Z.}~\bibnamefont {Jiang}}, \bibinfo {author} {\bibfnamefont
  {D.}~\bibnamefont {Kafri}}, \bibinfo {author} {\bibfnamefont
  {K.}~\bibnamefont {Kechedzhi}}, \bibinfo {author} {\bibfnamefont
  {J.}~\bibnamefont {Kelly}}, \bibinfo {author} {\bibfnamefont {P.~V.}\
  \bibnamefont {Klimov}}, \bibinfo {author} {\bibfnamefont {S.}~\bibnamefont
  {Knysh}}, \bibinfo {author} {\bibfnamefont {A.}~\bibnamefont {Korotkov}},
  \bibinfo {author} {\bibfnamefont {F.}~\bibnamefont {Kostritsa}}, \bibinfo
  {author} {\bibfnamefont {D.}~\bibnamefont {Landhuis}}, \bibinfo {author}
  {\bibfnamefont {M.}~\bibnamefont {Lindmark}}, \bibinfo {author}
  {\bibfnamefont {E.}~\bibnamefont {Lucero}}, \bibinfo {author} {\bibfnamefont
  {D.}~\bibnamefont {Lyakh}}, \bibinfo {author} {\bibfnamefont
  {S.}~\bibnamefont {Mandr\`a}}, \bibinfo {author} {\bibfnamefont {J.~R.}\
  \bibnamefont {McClean}}, \bibinfo {author} {\bibfnamefont {M.}~\bibnamefont
  {McEwen}}, \bibinfo {author} {\bibfnamefont {A.}~\bibnamefont {Megrant}},
  \bibinfo {author} {\bibfnamefont {X.}~\bibnamefont {Mi}}, \bibinfo {author}
  {\bibfnamefont {K.}~\bibnamefont {Michielsen}}, \bibinfo {author}
  {\bibfnamefont {M.}~\bibnamefont {Mohseni}}, \bibinfo {author} {\bibfnamefont
  {J.}~\bibnamefont {Mutus}}, \bibinfo {author} {\bibfnamefont
  {O.}~\bibnamefont {Naaman}}, \bibinfo {author} {\bibfnamefont
  {M.}~\bibnamefont {Neeley}}, \bibinfo {author} {\bibfnamefont
  {C.}~\bibnamefont {Neill}}, \bibinfo {author} {\bibfnamefont {M.~Y.}\
  \bibnamefont {Niu}}, \bibinfo {author} {\bibfnamefont {E.}~\bibnamefont
  {Ostby}}, \bibinfo {author} {\bibfnamefont {A.}~\bibnamefont {Petukhov}},
  \bibinfo {author} {\bibfnamefont {J.~C.}\ \bibnamefont {Platt}}, \bibinfo
  {author} {\bibfnamefont {C.}~\bibnamefont {Quintana}}, \bibinfo {author}
  {\bibfnamefont {E.~G.}\ \bibnamefont {Rieffel}}, \bibinfo {author}
  {\bibfnamefont {P.}~\bibnamefont {Roushan}}, \bibinfo {author} {\bibfnamefont
  {N.~C.}\ \bibnamefont {Rubin}}, \bibinfo {author} {\bibfnamefont
  {D.}~\bibnamefont {Sank}}, \bibinfo {author} {\bibfnamefont {K.~J.}\
  \bibnamefont {Satzinger}}, \bibinfo {author} {\bibfnamefont {V.}~\bibnamefont
  {Smelyanskiy}}, \bibinfo {author} {\bibfnamefont {K.~J.}\ \bibnamefont
  {Sung}}, \bibinfo {author} {\bibfnamefont {M.~D.}\ \bibnamefont
  {Trevithick}}, \bibinfo {author} {\bibfnamefont {A.}~\bibnamefont
  {Vainsencher}}, \bibinfo {author} {\bibfnamefont {B.}~\bibnamefont
  {Villalonga}}, \bibinfo {author} {\bibfnamefont {T.}~\bibnamefont {White}},
  \bibinfo {author} {\bibfnamefont {Z.~J.}\ \bibnamefont {Yao}}, \bibinfo
  {author} {\bibfnamefont {P.}~\bibnamefont {Yeh}}, \bibinfo {author}
  {\bibfnamefont {A.}~\bibnamefont {Zalcman}}, \bibinfo {author} {\bibfnamefont
  {H.}~\bibnamefont {Neven}},\ and\ \bibinfo {author} {\bibfnamefont {J.~M.}\
  \bibnamefont {Martinis}},\ }\bibfield  {title} {\bibinfo {title} {Quantum
  supremacy using a programmable superconducting processor},\ }\href@noop {}
  {\bibfield  {journal} {\bibinfo  {journal} {Nature}\ }\textbf {\bibinfo
  {volume} {574}},\ \bibinfo {pages} {505} (\bibinfo {year}
  {2019})}\BibitemShut {NoStop}%
\bibitem [{\citenamefont {Kadowaki}\ and\ \citenamefont
  {Nishimori}(1998)}]{Kadowaki:98}%
  \BibitemOpen
  \bibfield  {author} {\bibinfo {author} {\bibfnamefont {T.}~\bibnamefont
  {Kadowaki}}\ and\ \bibinfo {author} {\bibfnamefont {H.}~\bibnamefont
  {Nishimori}},\ }\bibfield  {title} {\bibinfo {title} {Quantum annealing in
  the transverse {I}sing model},\ }\href@noop {} {\bibfield  {journal}
  {\bibinfo  {journal} {Phys.\ Rev.\ E}\ }\textbf {\bibinfo {volume} {58}},\
  \bibinfo {pages} {5355} (\bibinfo {year} {1998})}\BibitemShut {NoStop}%
\bibitem [{\citenamefont {Brooke}\ \emph {et~al.}(1999)\citenamefont {Brooke},
  \citenamefont {Bitko}, \citenamefont {Rosenbaum},\ and\ \citenamefont
  {Aeppli}}]{Brooke:99}%
  \BibitemOpen
  \bibfield  {author} {\bibinfo {author} {\bibfnamefont {J.}~\bibnamefont
  {Brooke}}, \bibinfo {author} {\bibfnamefont {D.}~\bibnamefont {Bitko}},
  \bibinfo {author} {\bibfnamefont {T.~F.}\ \bibnamefont {Rosenbaum}},\ and\
  \bibinfo {author} {\bibfnamefont {G.}~\bibnamefont {Aeppli}},\ }\bibfield
  {title} {\bibinfo {title} {Quantum annealing of a disordered magnet},\
  }\href@noop {} {\bibfield  {journal} {\bibinfo  {journal} {Science}\ }\textbf
  {\bibinfo {volume} {284}},\ \bibinfo {pages} {779} (\bibinfo {year}
  {1999})}\BibitemShut {NoStop}%
\bibitem [{\citenamefont {Boixo}\ \emph {et~al.}(2014)\citenamefont {Boixo},
  \citenamefont {R{\o}nnow}, \citenamefont {Isakov}, \citenamefont {Wang},
  \citenamefont {Wecker}, \citenamefont {Lidar}, \citenamefont {Martinis},\
  and\ \citenamefont {Troyer}}]{Boixo:14}%
  \BibitemOpen
  \bibfield  {author} {\bibinfo {author} {\bibfnamefont {S.}~\bibnamefont
  {Boixo}}, \bibinfo {author} {\bibfnamefont {T.~F.}\ \bibnamefont
  {R{\o}nnow}}, \bibinfo {author} {\bibfnamefont {S.~V.}\ \bibnamefont
  {Isakov}}, \bibinfo {author} {\bibfnamefont {Z.}~\bibnamefont {Wang}},
  \bibinfo {author} {\bibfnamefont {D.}~\bibnamefont {Wecker}}, \bibinfo
  {author} {\bibfnamefont {D.~A.}\ \bibnamefont {Lidar}}, \bibinfo {author}
  {\bibfnamefont {J.~M.}\ \bibnamefont {Martinis}},\ and\ \bibinfo {author}
  {\bibfnamefont {M.}~\bibnamefont {Troyer}},\ }\bibfield  {title} {\bibinfo
  {title} {Evidence for quantum annealing with more than one hundred qubits},\
  }\href@noop {} {\bibfield  {journal} {\bibinfo  {journal} {Nat.\ Phys.}\
  }\textbf {\bibinfo {volume} {10}},\ \bibinfo {pages} {218} (\bibinfo {year}
  {2014})}\BibitemShut {NoStop}%
\bibitem [{\citenamefont {Kirkpatrick}\ \emph {et~al.}(1983)\citenamefont
  {Kirkpatrick}, \citenamefont {Gelatt},\ and\ \citenamefont
  {Vecchi}}]{Kirkpatrick:83}%
  \BibitemOpen
  \bibfield  {author} {\bibinfo {author} {\bibfnamefont {S.}~\bibnamefont
  {Kirkpatrick}}, \bibinfo {author} {\bibfnamefont {C.~D.}\ \bibnamefont
  {Gelatt}},\ and\ \bibinfo {author} {\bibfnamefont {M.~P.}\ \bibnamefont
  {Vecchi}},\ }\bibfield  {title} {\bibinfo {title} {Optimization by simulated
  annealing},\ }\href@noop {} {\bibfield  {journal} {\bibinfo  {journal}
  {Science}\ }\textbf {\bibinfo {volume} {220}},\ \bibinfo {pages} {671}
  (\bibinfo {year} {1983})}\BibitemShut {NoStop}%
\bibitem [{\citenamefont {Hopfield}\ and\ \citenamefont
  {Tank}(1985)}]{Hopfield:85}%
  \BibitemOpen
  \bibfield  {author} {\bibinfo {author} {\bibfnamefont {J.~J.}\ \bibnamefont
  {Hopfield}}\ and\ \bibinfo {author} {\bibfnamefont {D.~W.}\ \bibnamefont
  {Tank}},\ }\bibfield  {title} {\bibinfo {title} {Neural computation of
  decisions in optimization problems},\ }\href@noop {} {\bibfield  {journal}
  {\bibinfo  {journal} {Biol.\ Cybern.}\ }\textbf {\bibinfo {volume} {52}},\
  \bibinfo {pages} {141} (\bibinfo {year} {1985})}\BibitemShut {NoStop}%
\bibitem [{\citenamefont {Peterson}\ and\ \citenamefont
  {Anderson}(1988)}]{Peterson:88}%
  \BibitemOpen
  \bibfield  {author} {\bibinfo {author} {\bibfnamefont {C.}~\bibnamefont
  {Peterson}}\ and\ \bibinfo {author} {\bibfnamefont {J.~R.}\ \bibnamefont
  {Anderson}},\ }\bibfield  {title} {\bibinfo {title} {Neural networks and
  {NP}-complete problems; a performance study of the graph bisectioning
  problem},\ }\href@noop {} {\bibfield  {journal} {\bibinfo  {journal} {Complex
  Syst.}\ }\textbf {\bibinfo {volume} {2}},\ \bibinfo {pages} {59} (\bibinfo
  {year} {1988})}\BibitemShut {NoStop}%
\bibitem [{\citenamefont {Lucas}(2014)}]{Lucas:14}%
  \BibitemOpen
  \bibfield  {author} {\bibinfo {author} {\bibfnamefont {A.}~\bibnamefont
  {Lucas}},\ }\bibfield  {title} {\bibinfo {title} {Ising formulations of many
  {NP} problems},\ }\href@noop {} {\bibfield  {journal} {\bibinfo  {journal}
  {Front. Phys.}\ }\textbf {\bibinfo {volume} {2}},\ \bibinfo {pages} {1}
  (\bibinfo {year} {2014})}\BibitemShut {NoStop}%
\bibitem [{\citenamefont {Perdomo}\ \emph {et~al.}(2008)\citenamefont
  {Perdomo}, \citenamefont {Truncik}, \citenamefont {Tubert-Brohman},
  \citenamefont {Rose},\ and\ \citenamefont {Aspuru-Guzik}}]{Perdomo:08}%
  \BibitemOpen
  \bibfield  {author} {\bibinfo {author} {\bibfnamefont {A.}~\bibnamefont
  {Perdomo}}, \bibinfo {author} {\bibfnamefont {C.}~\bibnamefont {Truncik}},
  \bibinfo {author} {\bibfnamefont {I.}~\bibnamefont {Tubert-Brohman}},
  \bibinfo {author} {\bibfnamefont {G.}~\bibnamefont {Rose}},\ and\ \bibinfo
  {author} {\bibfnamefont {A.}~\bibnamefont {Aspuru-Guzik}},\ }\bibfield
  {title} {\bibinfo {title} {Construction of model hamiltonians for adiabatic
  quantum computation and its application to finding low-energy conformations
  of lattice protein models},\ }\href@noop {} {\bibfield  {journal} {\bibinfo
  {journal} {Phys.\ Rev.\ A}\ }\textbf {\bibinfo {volume} {78}},\ \bibinfo
  {pages} {012320} (\bibinfo {year} {2008})}\BibitemShut {NoStop}%
\bibitem [{\citenamefont {Lau}\ and\ \citenamefont {Dill}(1989)}]{Lau:89}%
  \BibitemOpen
  \bibfield  {author} {\bibinfo {author} {\bibfnamefont {K.~F.}\ \bibnamefont
  {Lau}}\ and\ \bibinfo {author} {\bibfnamefont {K.~A.}\ \bibnamefont {Dill}},\
  }\bibfield  {title} {\bibinfo {title} {A lattice statistical mechanics model
  of the conformational and sequence spaces of proteins},\ }\href@noop {}
  {\bibfield  {journal} {\bibinfo  {journal} {Macromolecules}\ }\textbf
  {\bibinfo {volume} {22}},\ \bibinfo {pages} {3986} (\bibinfo {year}
  {1989})}\BibitemShut {NoStop}%
\bibitem [{\citenamefont {Perdomo-Ortiz}\ \emph {et~al.}(2012)\citenamefont
  {Perdomo-Ortiz}, \citenamefont {Dickson}, \citenamefont {Drew-Brook},
  \citenamefont {Rose},\ and\ \citenamefont {Aspuru-Guzik}}]{Perdomo-Ortiz:12}%
  \BibitemOpen
  \bibfield  {author} {\bibinfo {author} {\bibfnamefont {A.}~\bibnamefont
  {Perdomo-Ortiz}}, \bibinfo {author} {\bibfnamefont {N.}~\bibnamefont
  {Dickson}}, \bibinfo {author} {\bibfnamefont {M.}~\bibnamefont {Drew-Brook}},
  \bibinfo {author} {\bibfnamefont {G.}~\bibnamefont {Rose}},\ and\ \bibinfo
  {author} {\bibfnamefont {A.}~\bibnamefont {Aspuru-Guzik}},\ }\bibfield
  {title} {\bibinfo {title} {Finding low-energy conformations of lattice
  protein models by quantum annealing},\ }\href@noop {} {\bibfield  {journal}
  {\bibinfo  {journal} {Sci. Rep.}\ }\textbf {\bibinfo {volume} {2}},\ \bibinfo
  {pages} {248} (\bibinfo {year} {2012})}\BibitemShut {NoStop}%
\bibitem [{\citenamefont {Robert}\ \emph {et~al.}(2021)\citenamefont {Robert},
  \citenamefont {Barkoutsos}, \citenamefont {Woerner},\ and\ \citenamefont
  {Tavernelli}}]{Robert:21}%
  \BibitemOpen
  \bibfield  {author} {\bibinfo {author} {\bibfnamefont {A.}~\bibnamefont
  {Robert}}, \bibinfo {author} {\bibfnamefont {P.~K.}\ \bibnamefont
  {Barkoutsos}}, \bibinfo {author} {\bibfnamefont {S.}~\bibnamefont
  {Woerner}},\ and\ \bibinfo {author} {\bibfnamefont {I.}~\bibnamefont
  {Tavernelli}},\ }\bibfield  {title} {\bibinfo {title} {Resource-efficient
  quantum algorithm for protein folding},\ }\href@noop {} {\bibfield  {journal}
  {\bibinfo  {journal} {Npj\ Quantum\ Inf.}\ }\textbf {\bibinfo {volume} {7}},\
  \bibinfo {pages} {38} (\bibinfo {year} {2021})}\BibitemShut {NoStop}%
\bibitem [{\citenamefont {Outeiral}\ \emph {et~al.}(2021)\citenamefont
  {Outeiral}, \citenamefont {Morris}, \citenamefont {Shi}, \citenamefont
  {Strahm}, \citenamefont {Benjamin},\ and\ \citenamefont
  {Deane}}]{Outeiral:21}%
  \BibitemOpen
  \bibfield  {author} {\bibinfo {author} {\bibfnamefont {C.}~\bibnamefont
  {Outeiral}}, \bibinfo {author} {\bibfnamefont {G.~M.}\ \bibnamefont
  {Morris}}, \bibinfo {author} {\bibfnamefont {J.}~\bibnamefont {Shi}},
  \bibinfo {author} {\bibfnamefont {M.}~\bibnamefont {Strahm}}, \bibinfo
  {author} {\bibfnamefont {S.~C.}\ \bibnamefont {Benjamin}},\ and\ \bibinfo
  {author} {\bibfnamefont {C.~M.}\ \bibnamefont {Deane}},\ }\bibfield  {title}
  {\bibinfo {title} {Investigating the potential for a limited quantum speedup
  on protein lattice problems},\ }\href@noop {} {\bibfield  {journal} {\bibinfo
   {journal} {New\ J.\ Phys.}\ }\textbf {\bibinfo {volume} {23}},\ \bibinfo
  {pages} {103030} (\bibinfo {year} {2021})}\BibitemShut {NoStop}%
\bibitem [{\citenamefont {Micheletti}\ \emph {et~al.}(2021)\citenamefont
  {Micheletti}, \citenamefont {Hauke},\ and\ \citenamefont
  {Faccioli}}]{Micheletti:21}%
  \BibitemOpen
  \bibfield  {author} {\bibinfo {author} {\bibfnamefont {C.}~\bibnamefont
  {Micheletti}}, \bibinfo {author} {\bibfnamefont {P.}~\bibnamefont {Hauke}},\
  and\ \bibinfo {author} {\bibfnamefont {P.}~\bibnamefont {Faccioli}},\
  }\bibfield  {title} {\bibinfo {title} {Polymer physics by quantum
  computing},\ }\href@noop {} {\bibfield  {journal} {\bibinfo  {journal}
  {Phys.\ Rev.\ Lett.}\ }\textbf {\bibinfo {volume} {127}},\ \bibinfo {pages}
  {080501} (\bibinfo {year} {2021})}\BibitemShut {NoStop}%
\bibitem [{\citenamefont {Babbush}\ \emph {et~al.}(2014)\citenamefont
  {Babbush}, \citenamefont {Perdomo-Ortiz}, \citenamefont {O'Gorman},
  \citenamefont {Macready},\ and\ \citenamefont {Aspuru-Guzik}}]{Babbush:14}%
  \BibitemOpen
  \bibfield  {author} {\bibinfo {author} {\bibfnamefont {R.}~\bibnamefont
  {Babbush}}, \bibinfo {author} {\bibfnamefont {A.}~\bibnamefont
  {Perdomo-Ortiz}}, \bibinfo {author} {\bibfnamefont {B.}~\bibnamefont
  {O'Gorman}}, \bibinfo {author} {\bibfnamefont {W.}~\bibnamefont {Macready}},\
  and\ \bibinfo {author} {\bibfnamefont {A.}~\bibnamefont {Aspuru-Guzik}},\
  }\bibfield  {title} {\bibinfo {title} {Construction of energy functions for
  lattice heteropolymer models: efficient encodings for constraint satisfaction
  programming and quantum annealing},\ }\href@noop {} {\bibfield  {journal}
  {\bibinfo  {journal} {Adv.\ Chem.\ Phys.}\ }\textbf {\bibinfo {volume}
  {155}},\ \bibinfo {pages} {201} (\bibinfo {year} {2014})}\BibitemShut
  {NoStop}%
\bibitem [{\citenamefont {Reva}\ and\ \citenamefont
  {Finkelstein}(2014)}]{Reva:96}%
  \BibitemOpen
  \bibfield  {author} {\bibinfo {author} {\bibfnamefont {B.~A.}\ \bibnamefont
  {Reva}}\ and\ \bibinfo {author} {\bibfnamefont {A.~V.}\ \bibnamefont
  {Finkelstein}},\ }\bibfield  {title} {\bibinfo {title} {Search for the most
  stable folds of protein chains: {II}.~{C}omputation of stable architectures
  of $\beta$-proteins using a self-consistent molecular field theory},\
  }\href@noop {} {\bibfield  {journal} {\bibinfo  {journal} {Protein\ Eng.\
  Des.\ Sel.}\ }\textbf {\bibinfo {volume} {9}},\ \bibinfo {pages} {399}
  (\bibinfo {year} {2014})}\BibitemShut {NoStop}%
\bibitem [{\citenamefont {Crescenzi}\ \emph {et~al.}(1998)\citenamefont
  {Crescenzi}, \citenamefont {Goldman}, \citenamefont {Papadimitriou},
  \citenamefont {Piccolboni},\ and\ \citenamefont {Yannakakis}}]{Crescenzi:98}%
  \BibitemOpen
  \bibfield  {author} {\bibinfo {author} {\bibfnamefont {P.}~\bibnamefont
  {Crescenzi}}, \bibinfo {author} {\bibfnamefont {D.}~\bibnamefont {Goldman}},
  \bibinfo {author} {\bibfnamefont {C.}~\bibnamefont {Papadimitriou}}, \bibinfo
  {author} {\bibfnamefont {A.}~\bibnamefont {Piccolboni}},\ and\ \bibinfo
  {author} {\bibfnamefont {M.}~\bibnamefont {Yannakakis}},\ }\bibfield  {title}
  {\bibinfo {title} {On the complexity of protein folding},\ }\href@noop {}
  {\bibfield  {journal} {\bibinfo  {journal} {J.\ Comput.\ Biol.}\ }\textbf
  {\bibinfo {volume} {5}},\ \bibinfo {pages} {423} (\bibinfo {year}
  {1998})}\BibitemShut {NoStop}%
\bibitem [{\citenamefont {Irb\"ack}\ and\ \citenamefont
  {Troein}(2002)}]{Irback:02}%
  \BibitemOpen
  \bibfield  {author} {\bibinfo {author} {\bibfnamefont {A.}~\bibnamefont
  {Irb\"ack}}\ and\ \bibinfo {author} {\bibfnamefont {C.}~\bibnamefont
  {Troein}},\ }\bibfield  {title} {\bibinfo {title} {Enumerating designing
  sequences in the {HP} model},\ }\href@noop {} {\bibfield  {journal} {\bibinfo
   {journal} {J.\ Biol.\ Phys.}\ }\textbf {\bibinfo {volume} {28}},\ \bibinfo
  {pages} {1} (\bibinfo {year} {2002})}\BibitemShut {NoStop}%
\bibitem [{\citenamefont {Holzgr\"afe}\ \emph {et~al.}(2011)\citenamefont
  {Holzgr\"afe}, \citenamefont {Irb\"ack},\ and\ \citenamefont
  {Troein}}]{Holzgrafe:11}%
  \BibitemOpen
  \bibfield  {author} {\bibinfo {author} {\bibfnamefont {C.}~\bibnamefont
  {Holzgr\"afe}}, \bibinfo {author} {\bibfnamefont {A.}~\bibnamefont
  {Irb\"ack}},\ and\ \bibinfo {author} {\bibfnamefont {C.}~\bibnamefont
  {Troein}},\ }\bibfield  {title} {\bibinfo {title} {Mutation-induced fold
  switching among lattice proteins},\ }\href@noop {} {\bibfield  {journal}
  {\bibinfo  {journal} {J.\ Chem.\ Phys.}\ }\textbf {\bibinfo {volume} {135}},\
  \bibinfo {pages} {195101} (\bibinfo {year} {2011})}\BibitemShut {NoStop}%
\bibitem [{\citenamefont {Unger}\ and\ \citenamefont {Moult}(1993)}]{Unger:93}%
  \BibitemOpen
  \bibfield  {author} {\bibinfo {author} {\bibfnamefont {R.}~\bibnamefont
  {Unger}}\ and\ \bibinfo {author} {\bibfnamefont {J.}~\bibnamefont {Moult}},\
  }\bibfield  {title} {\bibinfo {title} {Genetic algorithms for protein folding
  simulations},\ }\href@noop {} {\bibfield  {journal} {\bibinfo  {journal} {J.\
  Mol.\ Biol.}\ }\textbf {\bibinfo {volume} {231}},\ \bibinfo {pages} {75}
  (\bibinfo {year} {1993})}\BibitemShut {NoStop}%
\bibitem [{\citenamefont {Bastolla}\ \emph {et~al.}(1998)\citenamefont
  {Bastolla}, \citenamefont {Frauenkron}, \citenamefont {Gerstner},
  \citenamefont {Grassberger},\ and\ \citenamefont {Nadler}}]{Bastolla:98}%
  \BibitemOpen
  \bibfield  {author} {\bibinfo {author} {\bibfnamefont {U.}~\bibnamefont
  {Bastolla}}, \bibinfo {author} {\bibfnamefont {H.}~\bibnamefont
  {Frauenkron}}, \bibinfo {author} {\bibfnamefont {E.}~\bibnamefont
  {Gerstner}}, \bibinfo {author} {\bibfnamefont {P.}~\bibnamefont
  {Grassberger}},\ and\ \bibinfo {author} {\bibfnamefont {W.}~\bibnamefont
  {Nadler}},\ }\bibfield  {title} {\bibinfo {title} {Testing a new {M}onte
  {C}arlo algorithm for protein folding},\ }\href@noop {} {\bibfield  {journal}
  {\bibinfo  {journal} {Protein\ Eng.}\ }\textbf {\bibinfo {volume} {32}},\
  \bibinfo {pages} {52} (\bibinfo {year} {1998})}\BibitemShut {NoStop}%
\bibitem [{\citenamefont {Liang}\ and\ \citenamefont {Wong}(2001)}]{Liang:01}%
  \BibitemOpen
  \bibfield  {author} {\bibinfo {author} {\bibfnamefont {F.}~\bibnamefont
  {Liang}}\ and\ \bibinfo {author} {\bibfnamefont {W.~H.}\ \bibnamefont
  {Wong}},\ }\bibfield  {title} {\bibinfo {title} {Evolutionary {M}onte {C}arlo
  for protein folding simulations},\ }\href@noop {} {\bibfield  {journal}
  {\bibinfo  {journal} {J.\ Chem.\ Phys.}\ }\textbf {\bibinfo {volume} {115}},\
  \bibinfo {pages} {3374} (\bibinfo {year} {2001})}\BibitemShut {NoStop}%
\bibitem [{\citenamefont {McGeoch}\ and\ \citenamefont
  {Farr\'e}(2020)}]{McGeoch:20}%
  \BibitemOpen
  \bibfield  {author} {\bibinfo {author} {\bibfnamefont {C.}~\bibnamefont
  {McGeoch}}\ and\ \bibinfo {author} {\bibfnamefont {P.}~\bibnamefont
  {Farr\'e}},\ }\href@noop {} {\emph {\bibinfo {title} {The D-Wave Advantage
  System: an overview}}},\ \bibinfo {type} {Tech. Rep.}\ (\bibinfo
  {institution} {D-Wave Systems Inc.},\ \bibinfo {year} {2020})\BibitemShut
  {NoStop}%
\bibitem [{\citenamefont {McGeoch}\ \emph {et~al.}(2020)\citenamefont
  {McGeoch}, \citenamefont {Farr\'e},\ and\ \citenamefont
  {Bernoudy}}]{McGeoch:20b}%
  \BibitemOpen
  \bibfield  {author} {\bibinfo {author} {\bibfnamefont {C.}~\bibnamefont
  {McGeoch}}, \bibinfo {author} {\bibfnamefont {P.}~\bibnamefont {Farr\'e}},\
  and\ \bibinfo {author} {\bibfnamefont {W.}~\bibnamefont {Bernoudy}},\
  }\href@noop {} {\emph {\bibinfo {title} {D-Wave Hybrid Solver Service +
  Advantage: technology update}}},\ \bibinfo {type} {Tech. Rep.}\ (\bibinfo
  {institution} {D-Wave Systems Inc.},\ \bibinfo {year} {2020})\BibitemShut
  {NoStop}%
\bibitem [{\citenamefont {Pearson}\ \emph {et~al.}(2019)\citenamefont
  {Pearson}, \citenamefont {Mishra}, \citenamefont {Hen},\ and\ \citenamefont
  {Lidar}}]{Pearson:19}%
  \BibitemOpen
  \bibfield  {author} {\bibinfo {author} {\bibfnamefont {A.}~\bibnamefont
  {Pearson}}, \bibinfo {author} {\bibfnamefont {A.}~\bibnamefont {Mishra}},
  \bibinfo {author} {\bibfnamefont {I.}~\bibnamefont {Hen}},\ and\ \bibinfo
  {author} {\bibfnamefont {D.~A.}\ \bibnamefont {Lidar}},\ }\bibfield  {title}
  {\bibinfo {title} {Analog errors in quantum annealing: doom and hope},\
  }\href@noop {} {\bibfield  {journal} {\bibinfo  {journal} {Npj\ Quantum\
  Inf.}\ }\textbf {\bibinfo {volume} {5}},\ \bibinfo {pages} {107} (\bibinfo
  {year} {2019})}\BibitemShut {NoStop}%
\bibitem [{\citenamefont {Willsch}\ \emph {et~al.}(2022)\citenamefont
  {Willsch}, \citenamefont {Willsch}, \citenamefont {Calaza}, \citenamefont
  {Jin}, \citenamefont {De~Raedt}, \citenamefont {Svensson},\ and\
  \citenamefont {Michielsen}}]{Willsch:22}%
  \BibitemOpen
  \bibfield  {author} {\bibinfo {author} {\bibfnamefont {D.}~\bibnamefont
  {Willsch}}, \bibinfo {author} {\bibfnamefont {M.}~\bibnamefont {Willsch}},
  \bibinfo {author} {\bibfnamefont {C.~D.~G.}\ \bibnamefont {Calaza}}, \bibinfo
  {author} {\bibfnamefont {F.}~\bibnamefont {Jin}}, \bibinfo {author}
  {\bibfnamefont {H.}~\bibnamefont {De~Raedt}}, \bibinfo {author}
  {\bibfnamefont {M.}~\bibnamefont {Svensson}},\ and\ \bibinfo {author}
  {\bibfnamefont {K.}~\bibnamefont {Michielsen}},\ }\bibfield  {title}
  {\bibinfo {title} {{Benchmarking Advantage and D-Wave 2000Q quantum annealers
  with exact cover problems}},\ }\href@noop {} {\bibfield  {journal} {\bibinfo
  {journal} {Quantum\ Inf.\ Process.}\ }\textbf {\bibinfo {volume} {21}},\
  \bibinfo {pages} {141} (\bibinfo {year} {2022})}\BibitemShut {NoStop}%
\bibitem [{\citenamefont {Miyazawa}\ and\ \citenamefont
  {Jernigan}(1996)}]{Miyazawa:96}%
  \BibitemOpen
  \bibfield  {author} {\bibinfo {author} {\bibfnamefont {S.}~\bibnamefont
  {Miyazawa}}\ and\ \bibinfo {author} {\bibfnamefont {R.~L.}\ \bibnamefont
  {Jernigan}},\ }\bibfield  {title} {\bibinfo {title} {Residue-residue
  potentials with a favorable contact pair term and an unfavorable high packing
  density term, for simulation and threading},\ }\href@noop {} {\bibfield
  {journal} {\bibinfo  {journal} {J.\ Mol.\ Biol.}\ }\textbf {\bibinfo {volume}
  {256}},\ \bibinfo {pages} {623} (\bibinfo {year} {1996})}\BibitemShut
  {NoStop}%
\end{thebibliography}
%apsrev4-2.bst 2019-01-14 (MD) hand-edited version of apsrev4-1.bst
%Control: key (0)
%Control: author (8) initials jnrlst
%Control: editor formatted (1) identically to author
%Control: production of article title (0) allowed
%Control: page (0) single
%Control: year (1) truncated
%Control: production of eprint (0) enabled
%

\end{document}